\documentclass[journal]{IEEEtran}

\IEEEoverridecommandlockouts
\normalsize
\usepackage{amssymb}
\usepackage{verbatim}
\usepackage{bbold}
\usepackage{bbm}
\usepackage[cmex10]{amsmath}
\usepackage{stfloats}
\usepackage{graphicx}
\usepackage{multirow}
\usepackage{subfigure}
\usepackage{tabularx}
\usepackage{booktabs}
\usepackage{amsmath}
\usepackage{epsfig,epsf,color,balance,cite}
\usepackage{indentfirst}
\usepackage{mathrsfs}
\usepackage{float}
\usepackage{algorithm}
\usepackage{algorithmic}

\usepackage{dsfont}
\usepackage{color}
\usepackage{url}
\usepackage{color}
\usepackage{graphicx}
\usepackage{epstopdf}
\usepackage{enumerate}
\usepackage{cite}

\begin{document}
% paper title
%title{Joint Optimization of Trajectory and Power Allocation for Unmanned Surface Vehicle Assisted  Maritime Wireless Network }
\title{ AdapINT: A Flexible and Adaptive In-Band Network Telemetry System Based on Deep Reinforcement Learning
}

\author{
\IEEEauthorblockN{ Penghui Zhang,~\IEEEmembership {Student Member,~IEEE,}
                   Hua Zhang,~\IEEEmembership {Member,~IEEE,} Yibo Pi,~\IEEEmembership {Member,~IEEE,}
Zijian Cao,~\IEEEmembership {Student Member,~IEEE,} Jingyu Wang, {Member,~IEEE,} Jianxin Liao, {Member,~IEEE}
}\\
\thanks{Penghui Zhang, Hua Zhang, and Zijian Cao are with the National Mobile Communications Research Laboratory, Southeast University, Nanjing 211111, China. (email: phzhang@seu.edu.cn, huazhang@seu.edu.cn, caozijian@seu.edu.cn)}
\thanks{Yibo Pi is with the UM-SJTU Joint Institute, Shanghai Jiao Tong University, Shanghai, 200240, China. (e-mail: yibo.pi@sjtu.edu.cn)}
\thanks{Jingyu Wang and Jianxin Liao are with the State Key Laboratory of Networking and Switching Technology, Beijing University of Posts and Telecommunications, Beijing 100876, China. (email: wangjingyu@bupt.edu.cn, liaojx@bupt.edu.cn)}

}

\maketitle
\vspace{-2.5em}\vspace{-2.5em}
\begin{abstract}

In-band Network Telemetry (INT) has emerged as a promising network measurement technology. However, existing network telemetry systems lack the flexibility to meet diverse telemetry requirements and are also difficult to adapt to dynamic network environments.
In this paper, we propose AdapINT, a versatile and adaptive in-band network telemetry framework assisted by dual-timescale probes, including long-period auxiliary probes (APs) and short-period dynamic probes (DPs). 
Technically, the APs collect basic network status information, which is used for the path planning of DPs. 
To achieve full network coverage, we propose an auxiliary probes path deployment (APPD) algorithm based on the Depth-First-Search (DFS).
The DPs collect specific network information for telemetry tasks.
To ensure that the DPs can meet diverse telemetry requirements and adapt to dynamic network environments, we apply the deep reinforcement learning (DRL) technique and transfer learning method to design the dynamic probes path deployment (DPPD) algorithm. 
The evaluation results show that AdapINT can redesign the telemetry system according to telemetry requirements and network environments. AdapINT can reduce telemetry latency by 75\% in online games and video conferencing scenarios. For overhead-aware networks, AdapINT can reduce control overheads by 34\% in cloud computing services.

\end{abstract}
% key words
%\vspace{-1.5em}

\footnotetext{This work was supported by the National Key Research and Development Program of China under Grant(2020YFB1807803) %, Science and Technology Project of State Grid Corporation (Research on 5G Power Sevice Access Capability Evaluation and Planning Simulation Technology).
.}

\IEEEpeerreviewmaketitle
\vspace{-1em}
\section{Introduction}
\label{section1}
Network telemetry is important for various network management applications \cite{1}, such as fault location \cite{6}, congestion control \cite{7}, path verification \cite{8}, and more\cite{9292999}. 
Real-time and reliable network information is critical to comprehending the network status. 
However, designing efficient network telemetry systems remains challenging due to two key difficulties. 
First, the network telemetry system needs to have flexibility to meet various telemetry requirements.
Second, the network telemetry system should be adaptable to the dynamic network environment.
These two difficulties are summarized as follows.

\begin{itemize}
\item {\bf Flexibility.} With the development of network function virtualization (NFV) \cite{BHAMARE20171} and service function chaining (SFC) \cite{BHAMARE2016138}, 
there is a growing diversity of network applications with unique telemetry requirements \cite{10063936}, e.g., high accuracy, low control overhead, low latency, and full network coverage.
If a network telemetry system fails to meet the requirements of these applications, it is hard to provide personalized services for users, negatively impacting its effectiveness. Therefore, flexibility is a crucial characteristic of a network telemetry system.

\item {\bf Adaptability.} 
Dynamic network environments \cite{3} are characterized by frequent fluctuations in network load \cite{4} and the potential for link disconnections or congestion \cite{2}.
To maintain stability and reliability in dynamic network environments, the telemetry systems should have the adaptability to adjust the system parameters in time according to the changes in the network environment and optimize the quality and efficiency of the network telemetry \cite{9784426}.

\end{itemize}

Despite their advantages in terms of network visibility, scalability, and accuracy \cite{11}, existing telemetry systems, including both passive network telemetry (PNT) systems \cite{8897503, 9, 10, 18} and active network telemetry (ANT) systems \cite{12, 13, 14}, face significant challenges in overcoming the difficulties (discussing in Section \ref{section2}). 
One major issue is they are typically designed based on fixed telemetry requirements and cannot flexibly meet new requirements, which can limit their ability to respond to changing conditions or evolving demands.
Moreover, existing telemetry systems are often deployed for specific network loads and topologies, making them less adaptable to unexpected scenarios such as node failures or link disconnections. When such faults occur, the performance of these systems may suffer significant degradation, leading to potential disruptions and outages. To address these challenges, researchers need to develop innovative approaches that can improve flexibility and adaptability.

This paper proposes AdapINT, a novel in-band network telemetry framework that leverages dual-timescale probes to provide flexible and adaptable telemetry services. AdapINT comprises two types of probes: long-period auxiliary probes (APs) and short-period dynamic probes (DPs). The APs collect basic network status information, while the DPs gather specific network information required by applications.
To improve adaptability, AdapINT uses the basic network status collected by the APs to periodically update the DPs' forwarding paths. Specifically, during each long period, AdapINT injects APs into the network using source routing technology \cite{19} to collect basic network status such as the topology and load of the entire network. Based on the latest APs' telemetry reports, AdapINT updates the DPs’ forwarding paths to adapt to the dynamic network environment and uses DPs to collect specific network information required by network applications during each short period. 

The probe path planning algorithms of APs and DPs are crucial for ensuring system performance. Since the APs’ forwarding paths should cover the entire network to collect comprehensive basic network status information. We propose an auxiliary probes path deployment (APPD) algorithm based on the Depth-First-Search (DFS) algorithm to achieve full network coverage. To improve flexibility, AdapINT should be capable of meeting different telemetry requirements through the DPs' telemetry process. To achieve this, we propose a dynamic probes path deployment (DPPD) algorithm based on deep reinforcement learning (DRL). By adjusting the reward function, the DPPD algorithm can meet various telemetry requirements, such as shortest path, minimum bandwidth consumption, and minimum latency. 
Inspired by \cite{26}, our proposed DRL model consists of a recursive neural network (RNN) decoder and an attention mechanism. To reduce the complexity of the proposed DRL model, we require the RNN model to keep the input sequence unchanged, so that changing the order of any two inputs does not affect the network. Additionally, we set a mask to shield infeasible solutions, which not only avoids the routing loop problem but also facilitates faster model training. 
To adapt the DPPD algorithm to different telemetry requirements and environments, we introduce transfer learning for various reward functions and transfer learning for different training sets \cite{TL}, which can reduce the training time of the DRL model.

The contributions of this paper can be summarized as follows:

\vspace{-0em}
\begin{itemize}
\item We propose a flexible and adaptable in-band network telemetry framework, called AdapINT. It uses long-period APs to collect basic network status information and short-period DPs to collect various network information required by network applications, thus achieving on-demand, network-wide, stable network telemetry.

\item The DFS-based path planning algorithm is proposed for APs, called the APPD algorithm. The proposed algorithm can achieve full network coverage with low complexity.

\item The DPPD algorithm is proposed for DPs based on a self-learning DRL model. The proposed algorithm can meet various telemetry tasks by adjusting the reward and can constantly update the forwarding path according to real-time network information. Moreover, a transfer learning method is introduced to reduce the training time of the DRL model.

\item We develop a self-learning DRL model based on the attention mechanism. Considering the characteristics of the routing path planning problem, we improve the input queue of the RNN model and set a mask function to reduce model complexity.

%\item We propose a transfer learning method to accelerate adaptation to new telemetry requirements. The DPPD algorithm can reduce the training time of the DRL model and complete various telemetry tasks more efficiently.
\item We demonstrate the flexibility and adaptability of AdapINT under a variety of different network environments and telemetry requirements. Compared with traditional network telemetry systems, AdapINT can reduce telemetry latency by 75\% in online games and video conferencing scenarios and can reduce control overheads by 34\% in cloud computing services.
\end{itemize}

The rest of this paper is organized as follows.
In Section \ref{section2}, we discuss the related work.
Later, we propose an architecture of AdapINT and describe the critical designs for AdapINT in Section \ref{section3} and we propose the APPD algorithm in Section \ref{section4}. In Section \ref{section5}, we propose a DRL model and apply it to the design of the DPPD algorithm.
In Section \ref{section6}, simulation results are presented to demonstrate the performance of AdapINT. Finally, we conclude the paper in Section \ref{section7}.

\vspace{-1.0em}
\section{Related Work}
\label{section2}
This section discusses the related work, including PNT, ANT, and DRL.
\vspace{-1.0em}
\subsection{Passive Network Telemetry}
In-band network telemetry (INT) is an emerging representative of PNT technology due to the development of software-defined networks (SDN) \cite{SDN} and programmable data-plane technology \cite{PDP}. INT enables network devices to add device-internal information to packets, such as port numbers, link utilization, or queuing latency, which is then sent to the centralized controller for further analysis at the last hop. INT increases the packet size, which incurs additional overhead, and any methods have been proposed to reduce the telemetry overhead, such as PRoML-INT, PINT, and INT-label \cite{9, 10, 18}. 
However, due to the uncontrolled probe paths, PNT makes it difficult to achieve full network coverage and can result in the switches being repeatedly probed. The redundant information caused by repeated collection leads to a variety of problems, including increased difficulty of data analysis and higher computational overhead \cite{ML1}.

\vspace{-1.0em}
\subsection{Active Network Telemetry}
The basic idea of ANT \cite{28} is to generate a telemetry probe from an INT agent to probe the user-specified telemetry path actively \cite{19} through source routing technology. 
However, existing ANT systems only use flexible probe path planning algorithms to meet fixed telemetry requirements in different network scenarios, including INT-path, NetView, and IntOpt \cite{12, 13, 14}.
INT-path is designed to cover the entire network and reduce telemetry overhead by using the Depth-First-Search (DFS) algorithm and Euler Trail/Circuit-based path planning policy \cite{12}. NetView is designed to support multiple telemetry applications at different frequencies. It requires only one vantage server for the entire telemetry process and can cover the entire network topology \cite{13}. 
IntOpt is specifically designed for VNF service chain network monitoring and uses a simulated annealing-based random greedy metaheuristic (SARG) to minimize overhead during active probing and collection \cite{14}. 
These ANT systems tend to optimize specific performance metrics, such as latency or overhead, with fixed control strategies in limited types of scenarios.
They lack the flexibility to reconfigure and redeploy probe paths for new telemetry requirements \cite{5}. Furthermore, existing network telemetry systems are typically deployed based on fixed network environments and are hard to self-adjust when the network environment, such as network topology and network load changes. These environmental shifts are relatively common in real-world networks and can significantly impact the effectiveness of telemetry systems, requiring path planning strategies to be continuously adjusted to match the current network state.

\vspace{-1.2em}
\subsection{Deep Reinforcement Learning and Transfer Learning}
DRL provides new opportunities for solving multi-objective optimization and dynamic environment adaptation problems\cite{20, 21, 23, 24, 25}.
For example, DRL has significant potential in dealing with custom optimization objective problems\cite{9040280}.
It can continually adjust model parameters to adapt to various optimization objectives and efficiently meet new telemetry requirements without any human assistance\cite{16}. Therefore, we use it to enable network telemetry systems to have the ability to meet different telemetry requirements. 
Furthermore, DRL has excellent scalability and can solve problems with similar structures \cite{15}, making it ideal for updating path planning based on real-time network information \cite{ML2}. By leveraging transfer learning techniques, we can further reduce the training time of the DRL model by transferring knowledge from existing telemetry tasks to new ones \cite{TL}. This approach is crucial for enabling telemetry systems to adapt to evolving network environments and meet diverse telemetry requirements effectively.

\vspace{-0.8em}
\section{System Design}
\label{section3}
In this section, we first present the architecture of AdapINT. Then, we describe the critical designs for AdapINT, including probe format, switch behaviors, metadata, and queries. At last, we introduce how the controller handles the telemetry results.
\vspace{-1.em}

\subsection{Architecture of AdapINT}
\begin{figure}
\centering
\setlength{\abovecaptionskip}{0.cm}
\setlength{\belowcaptionskip}{-10 cm}
\includegraphics[width=10cm]{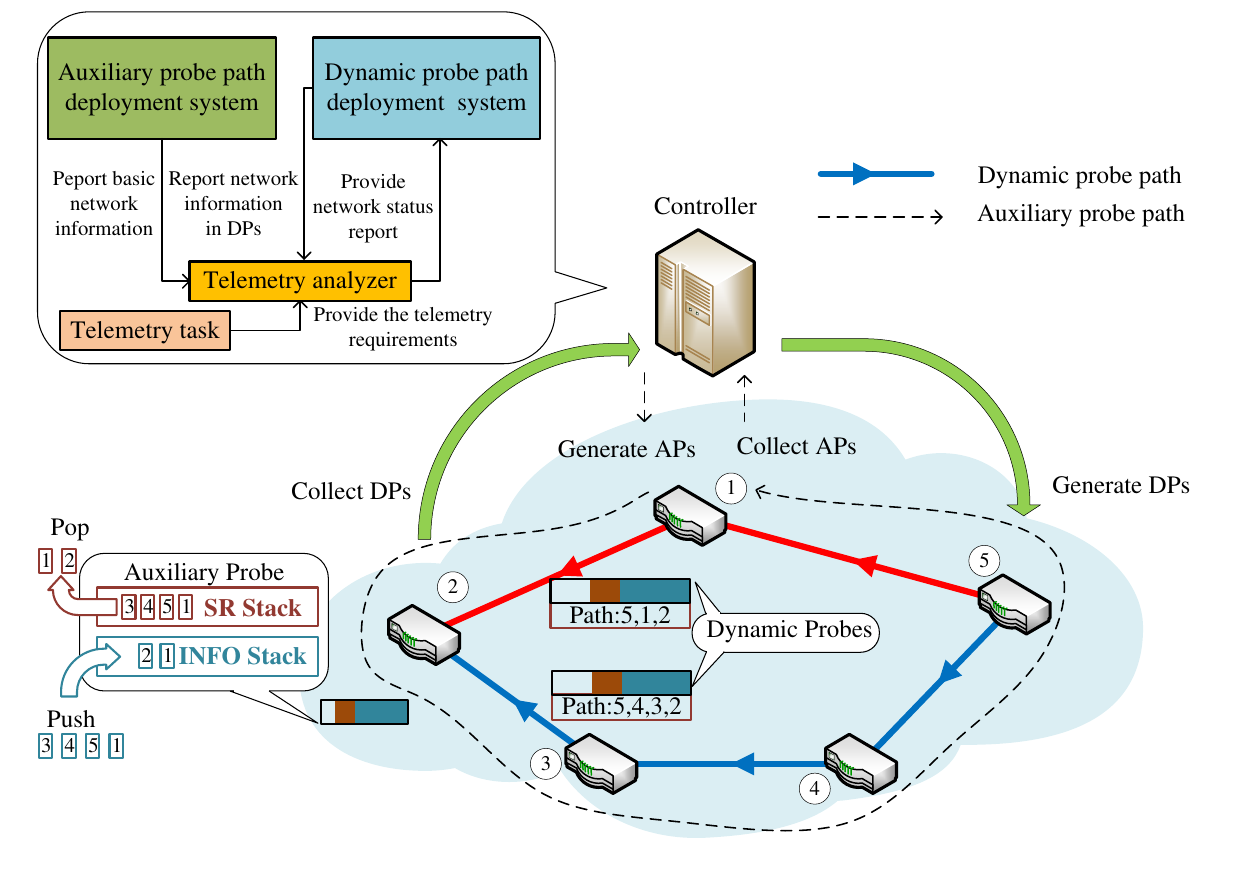}
\caption{Architecture of AdapINT.}
\label{fig1}
\vspace{-0.1cm}
\end{figure}

\begin{figure}
\centering
\setlength{\abovecaptionskip}{0.cm}
\setlength{\belowcaptionskip}{-10 cm}
\includegraphics[width=10cm]{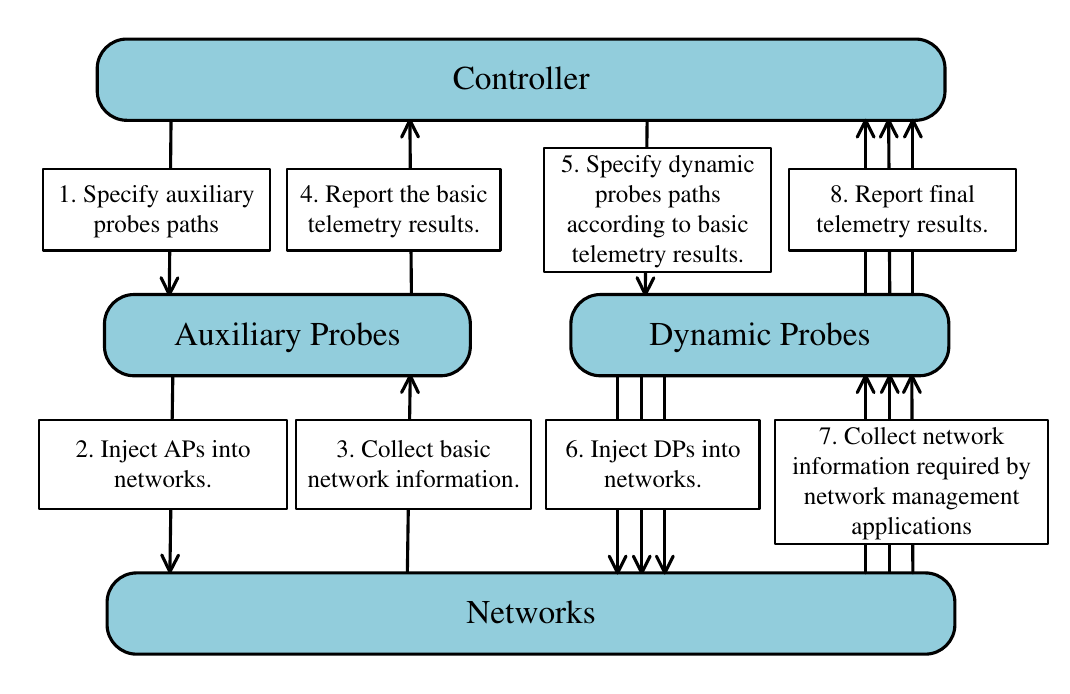}
\caption{Workflow of AdapINT.}
\label{fig1.2}
\vspace{-0.1cm}
\end{figure}

AdapINT is committed to designing an on-demand network telemetry system in a dynamic network environment. To improve the adaptability, AdapINT designs a network telemetry architecture comprised of dual-timescale probes \cite{12}.

Fig. \ref{fig1} illustrates that AdapINT comprises four main components: the auxiliary probe path deployment system, the dynamic probe path deployment system, the telemetry analyzer, and the telemetry task. Users can specify telemetry tasks through network applications to provide telemetry requirements to the telemetry analyzer. 
The auxiliary probe path deployment system designs auxiliary probe paths for APs to acquire basic network status, which is used by the dynamic probe path deployment system to deploy dynamic probe paths \cite{28}.
After receiving APs each time, the dynamic probe paths are updated based on the real-time network information. Besides, DPs are responsible for collecting network information required for network applications based on telemetry tasks at a higher frequency. Compared to APs, DPs take up more network resources and incur most of the telemetry overhead due to the higher frequency and more network information collected.

As shown in Fig. \ref{fig1.2}, the workflow of AdapINT can be described as follows: First, the controller plans the auxiliary probe paths according to the network topology structure. Second, the network device generates APs and injects them into the network at a low frequency. Third, APs are forwarded along auxiliary probe paths and collect basic network status. Next, after receiving APs, the controller obtains the basic telemetry results. Then, the controller designs dynamic probe paths based on real-time network information. Moreover, DPs are injected into the network at a high frequency. Furthermore, the DPs collect all required network information. At last, DPs report the collected network information to the controller for further analysis and processing.

\subsection{Probe Format}

\begin{figure*}
\centering
\setlength{\abovecaptionskip}{0.cm}
\setlength{\belowcaptionskip}{-10 cm}
\includegraphics[width=14cm]{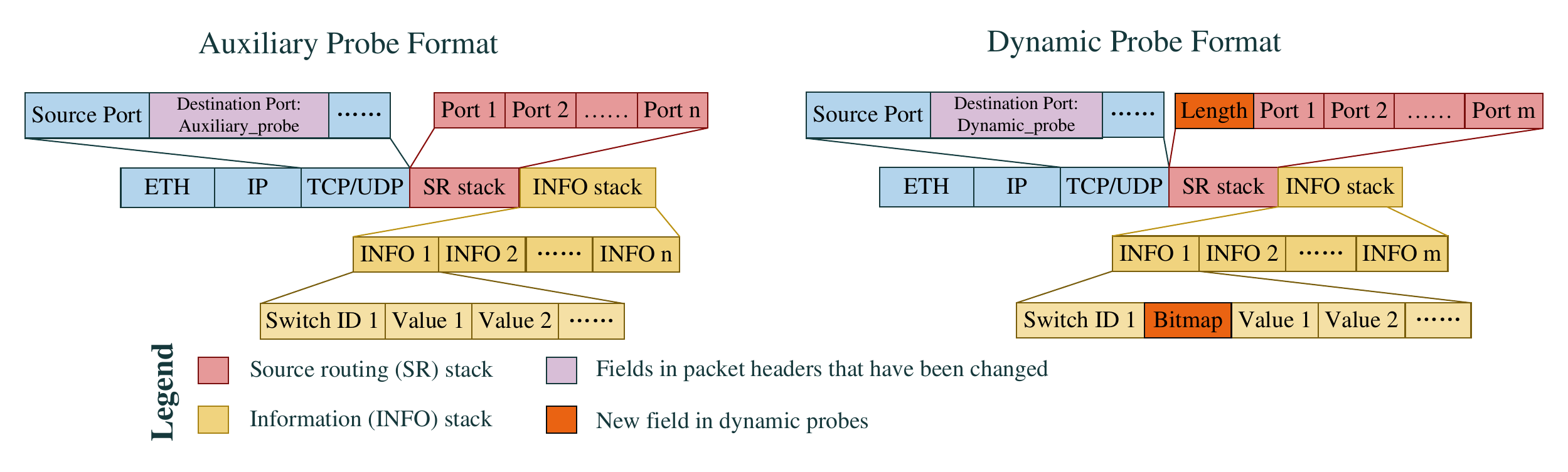}
\caption{Auxiliary probe format and dynamic probe format. }
\label{fig2}
\vspace{-0.1cm}
\end{figure*}
APs are responsible for collecting basic network status information, while DPs need to collect specific network information required by applications. Therefore, AdapINT designs the auxiliary probe format and dynamic probe format based on the characteristics of different tasks.

\subsubsection{Auxiliary probe format}
As shown in Fig. \ref{fig2}, the AP's header is similar to a normal packet header. It consists of an Ethernet header, an IP header, and a TCP/UDP header. The destination port is set to ``Auxiliary\_probe" to ensure that programmable switches can parse APs. In addition, the AP's destination address is set as the controller's IP to forward APs to the controller.

Moreover, the AP consists of a source routing stack (SR stack) and an information stack (INFO stack). The SR stack is a fixed-length stack consisting of a series of Port labels. Each Port label records one hop on the auxiliary probe path, and the switches forward APs based on the Port labels. The INFO stack is used to store network information. Each INFO label contains the switch ID and the value of each network metadata. 

\subsubsection{Dynamic probe format}
The dynamic probe format is similar to the auxiliary probe format. 
However, as the length of dynamic probe paths varies more widely, a fixed-length SR stack can result in unnecessary waste. Furthermore, given the diverse types of network metadata collected by DPs, switches need to identify which kind of metadata is to be collected.
Therefore, the dynamic probe format introduces two new fields: (i) Length and (ii) Bitmap.
As shown in Fig. \ref{fig2}, the SR stack is a variable length stack in the dynamic probe format, and the ``Length” determines the length of the SR stack. And the ``Bitmap" determines what kinds of metadata should be collected by the INFO label. It is worth mentioning that the destination port in the dynamic probe is set to ``Dynamic\_probe” so that the switches can parse DPs.

\vspace{-0.5cm}
\subsection{Switch Behaviours}

\begin{figure}
\centering
\setlength{\abovecaptionskip}{0.cm}
\setlength{\belowcaptionskip}{-10 cm}
\includegraphics[width=8.5cm]{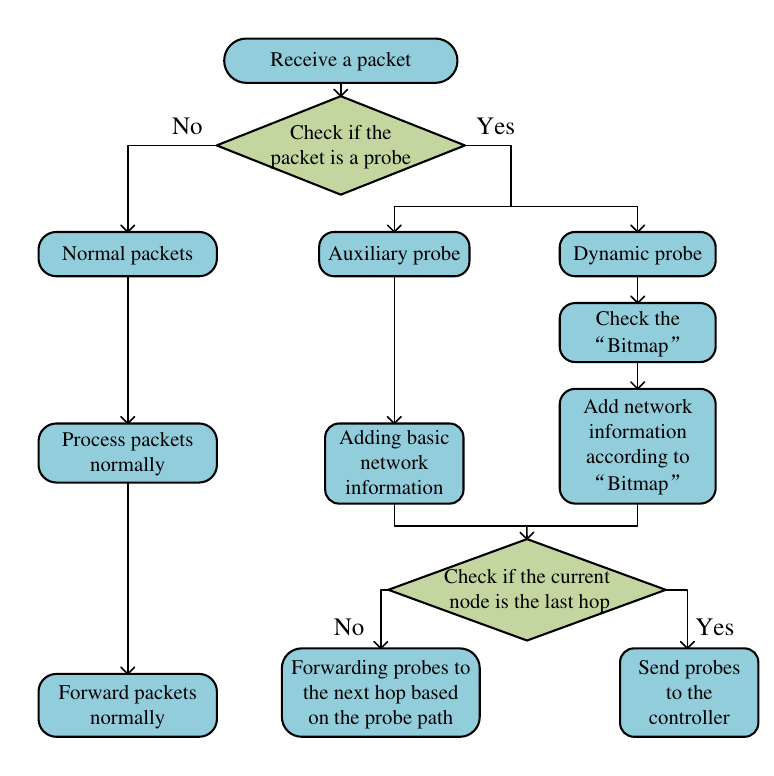}
\caption{Switch behaviours. }
\label{fig3}
\vspace{-0.1cm}
\end{figure}

The tasks of programmable switches  \cite{PDP} are recognizing, processing, and forwarding probes.
After receiving the packet, the switch behaviors are shown in Fig. \ref{fig3}.
Firstly, the switch needs to check if the packet is a probe. If the destination port is ``Auxiliary\_probe'' or ``Dynamic\_probe'', the switch considers this packet to be a probe. If it is not, the packet is processed and forwarded normally. Next, the switch should process the probe according to the probe type. For an AP, the switch should record the basic network information with the INFO label and add the INFO label to the INFO stack. For a DP,  the switch should add network information to the INFO label according to the ``Bitmap''. Finally, the switch at the last hop in the probe path sends the telemetry information to the controller for further analysis. Otherwise, the switch forwards the probe to the next hop according to the probe path.

\vspace{-0.5cm}
\subsection{Metadata and Queries}
Metadata represents the various network information values collected by AdapINT from switches. To simplify user operations, AdapINT employs queries to specify the types of metadata required for various telemetry tasks.

AdapINT can not only obtain metadata directly from switches, ingresses, egresses, and buffers, but also obtain high-level metadata through the calculation of the underlying data. In summary, AdapINT supports two types of metadata, namely node-level metadata and path-level metadata. Node-level metadata includes switch ID, switch workload, queue length, congestion status, queue packet loss rate, port packet loss rate, port timestamp, port packet count, and so on. Path-level metadata includes path latency, path utilization, path throughput, etc.

The user can write queries to represent one or multiple telemetry tasks. We use the following telemetry tasks as examples:
(a) Congestion Control. Users can let AdapINT periodically measure the ratio of the current queue length to the configured maximum queue limit to obtain the congestion status.
(b) Network fault diagnosis. After obtaining the link latency, users can quickly know the location of the network fault by determining whether the latency is abnormal or not.
Note that APs have a fixed telemetry task to obtain the network information needed for the DPPD algorithm. Therefore, APs have a unique query. In this paper, APs provide link latency and port status information for the DPPD algorithm.
\vspace{-0.5cm}
\subsection{Telemetry Results Analysis}
\label{section3F}
When probes are collected by the controller, the telemetry analyzer parses each probe into a series of dictionaries or tuples.
For APs, the analyzer provides basic network status to the dynamic probe path deployment system.
For DPs, the analyzer stores the network information in a database, which other network applications can access. 
To ensure structured storage, the storage format is designed according to the switch ID and metadata types. In order to save storage space, whenever new probes are received, the telemetry analyzer updates the contents of the existing dictionaries or tuples with the latest network information.
Finally, the telemetry analyzer calculates high-level metadata based on the queries. Network applications can periodically obtain network telemetry reports based on their telemetry requirements.

\begin{figure}
\centering
\setlength{\abovecaptionskip}{0.cm}
\setlength{\belowcaptionskip}{-10 cm}
\includegraphics[width=9cm]{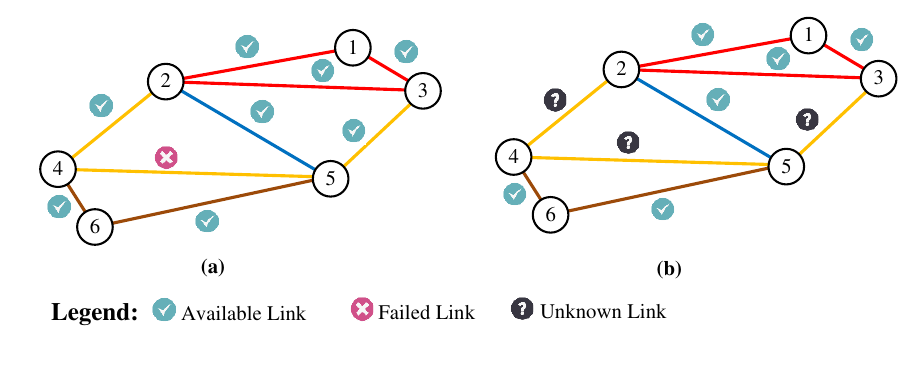}
\caption{Telemetry results analysis. }
\label{fig6}
\vspace{-0.1cm}
\end{figure}

Another task of the telemetry analyzer is to locate network faults via APs. 
When a link fails, all APs that pass through the faulty link cannot be collected by the controller. Therefore, it is necessary to check all the links that these APs pass through to locate the fault.
To minimize the number of APs that need to be checked, we must guarantee that only one AP passes through every link. By utilizing non-overlapping auxiliary probe paths, we can quickly find the path where the fault is located and reduce the workload of the fault location.
As illustrated in Fig. \ref{fig6}, if link $\left(4,5\right)$ is faulty, the APs with the paths $\left[ 2,4,5,3 \right]$ will not be received by the telemetry analyzer. Consequently, when locating the faulty link, the controller can narrow down the potential location to links $\left(2,4\right)$, $\left(4,5\right)$, and $\left(5,3\right)$.

\vspace{-1em}
\section{Auxiliary Probe Path Deployment System}
\label{section4}
In this section, we first analyze the problem of auxiliary probe path deployment. Then, we design the APPD algorithm based on DFS.
\vspace{-1em}
\subsection{Problem Analysis}
\label{section4A}

Considering a network containing $n$ routers, we define the network topology as an undirected physical graph, denoted as $G=\left( V, E \right)$. $V$ is the set of physical nodes denoted as $V=\left\{ \left. i \right|i=1,\cdots ,n \right\}$ and $E$ is the set of physical links denoted as $E=\left\{ \left. \left( i,j \right) \right|i,j\in V \right\}$. The index of the physical node is denoted by $i\in V$, and the physical links between node $i$ and node $j$ are denoted by $\left( i,j \right)$ or $\left( j, i \right)$. $E$ is the set of unordered binary groups consisting of the elements in $V$.

We denote the $p$-th auxiliary probe path as ${{{a}}_{p}}=\left[ {{v}_{p,1}},{{v}_{p,2}},\cdots ,{{v}_{p,N_{p}}} \right], p=1,2,\cdots , P$, where $N_{p}$ is the number of nodes in path $a_p$ passes and ${v}_{p,i}$ denotes the $i$-th node that path $a_p$ passes through. Since the SR stack of the auxiliary probe is a fixed-length stack, the length of the port information cannot exceed the length of the SR stack, which can be represented as
\begin{IEEEeqnarray}{rCl} %��ʽ3
\label{form_4.c}
{\!}
\begin{split}
\left| {{{a}}_{p}} \right|\le {{l}_{th}},\\
\end{split}
\end{IEEEeqnarray}
where $\left| {{{a}}_{p}} \right|$ denotes the length of the $p$-th path and ${l}_{th}$ denotes the length threshold of auxiliary probe paths. 

Comprehensive network status information is crucial for the path planning of DPs. As a result, it is crucial to ensure that APs can traverse all physical links within the network. This requirement can be described as follows:

\begin{IEEEeqnarray}{rCl} %��ʽ1
\label{form_4.a}
{\!}
\begin{split}
 E=\bigcup\limits_{p=1}^{P}{{{L}_{p}}},\\
\end{split}
\end{IEEEeqnarray}
where ${L}_{p}$ is the set of links which the $p$-th path passes through.

Moreover, to reduce the number of links that require verification by the controller during fault localization (as explained in Section \ref{section3F}), it is necessary to ensure that each link is probed by only one probe. The above requirement can be described as 
\begin{IEEEeqnarray}{rCl} %��ʽ2
\label{form_4.b}
{\!}
\begin{split}
{{L}_{{{p}_{1}}}}\bigcap{{{L}_{{{p}_{2}}}}=\varnothing,\text{  }\forall {{p}_{1}},{{p}_{2}}}\in \left\{ 1,2,\cdots ,P \right\},\text{  }{{p}_{1}}\ne {{p}_{2}},\\
\end{split}
\end{IEEEeqnarray}
where ${p}_{1}$, ${p}_{2}$ are the index of any two auxiliary probe paths. 

Since APs have lower frequency and less data carryover compared to DPs, their impact on network performance is minor. Thus, we consider the auxiliary probe path planning problem as finding feasible solutions that meet the following constraints: First, the length of each auxiliary probe path must not exceed the SR stack capacity (Constraint \eqref{form_4.c}). Second, the auxiliary probe path should cover the entire network topology (Constraint \eqref{form_4.a}). Finally, the auxiliary probe paths should be designed to avoid overlap with one another (Constraint \eqref{form_4.b}).

\vspace{-0.5em}
\subsection{Auxiliary Probes Path Deployment (APPD) Algorithm}
In this subsection, based on the analysis presented in Section \ref{section4A}, we propose an APPD algorithm to plan auxiliary probe paths and describe its operation steps in detail.

Once an auxiliary probe path is created, we should select and add suitable nodes to the path.
How to choose a suitable node is very important to improve the efficiency of path planning. Therefore, different rules for adding nodes will directly affect the performance of the APPD algorithm. Due to the Constraint \eqref{form_4.a} and Constraint \eqref{form_4.b}, we can propose a simple DFS-based algorithm to select the next node, which can generate non-overlapping paths to cover the whole network.
However, because the basic idea of the DFS algorithm is to explore the graph along each branch as much as possible, it may generate excessively long probe paths. To ensure that the length of the auxiliary probe path does not exceed the capacity of the SR stack, we must limit the depth of each exploration.

Let us assume that we are planning the $p$-th auxiliary probe path ${{{a}}_{p}},p=1,2,\cdots , P$. After adding the node ${{v}_{p,k-1}},k=2,\cdots ,{{l}_{th}}$, the $p$-th path can be temporarily denoted as ${{{a}}_{p}}=\left[ {{v}_{p,1}},{{v}_{p,2}},\cdots ,{{v}_{p,k-1}} \right]$. Next, we should look for the appropriate node ${v}_{p,k}$, which is the node connected to node ${v}_{p,k-1}$. The set of nodes connected to node ${v}_{p,k-1}$ can be denoted as 
\begin{IEEEeqnarray}{rCl} %��ʽ7
\label{form_7}
{\!}
\begin{split}
V\left( {{v}_{p,k-1}} \right)=\left\{ \left. i \right|\exists \left( {{v}_{p,k-1}},i \right)\in E \right\},\\
\end{split}
\end{IEEEeqnarray}
where the symbol $\exists$ indicates that the link is in the set $E$. We should choose node ${v}_{p,k}$ in such a way that the link $\left( {{v}_{p,k-1}},{v}_{p,k} \right)$ is not yet covered by existing auxiliary probe paths, which can be denoted as 
\begin{IEEEeqnarray}{rCl} %��ʽ8
\label{form_8}
{\!}
\begin{split}
\left( {{v}_{p,i-1}},{{v}_{p,i}} \right)\notin {{\bar{L}}},\\
\end{split}
\end{IEEEeqnarray}
where ${{\bar{L}}}=\bigcup\limits_{i=1}^{p}{{{L}_{i}}}$ is the set of links covered by existing auxiliary probe paths. Moreover, considering the constraint \eqref{form_4.c}, we also need to ensure that the length of path ${{a}}_{p}$ does not exceed $L_{th}$ after adding node ${v}_{p,k}$, which can be denoted as 
\begin{IEEEeqnarray}{rCl} %��ʽ9
\label{form_9}
{\!}
\begin{split}
\left| {{{{a}}_{p}}} \right|+1\le {{l}_{th}}.\\
\end{split}
\end{IEEEeqnarray}
Therefore, based on the DFS algorithm, the node selection rule of our proposed long-time-scale algorithm can be denoted as 
\begin{IEEEeqnarray}{rCl} %��ʽ10
\label{form_10}
{\!}
\begin{split}
{{v}_{p,k}}=\underset{i}{\mathop{\min }}\,i,i\in V\left( {{v}_{p,k-1}} \right).\\
\end{split}
\end{IEEEeqnarray}

Note that if we are unable to append node ${v}_{p,k}$ to path ${{a}}_{p}$ due to Eq. \eqref{form_9}, it is necessary to create a new path ${{a}_{p+1}}$ from node ${v_{p,k-1}}$ and start planning from that point onwards. In the event that no nodes satisfying Eq. \eqref{form_8} are found, we should go back to the previous node, which has the remaining unvisited neighbors, before creating a new path ${{a}}_{p+1}$.

To enhance clarity and facilitate description, we present the proposed APPD algorithm in Algorithm \ref{alg1}, followed by a detailed description of its specific steps:
\begin{enumerate}[{~~~~}Step 1)]
\item {\textit{Initialization:}}The algorithm parameters are initialized as follows: $l_{th}$ represents the length threshold of auxiliary probe paths, $v$ represents the node index, $n$ represents the length of paths, and $flag$ is used to control the creation of a new path.
\item {\textit{Traverse:}} First, select a node in the network and create a path. Following selection rule \eqref{form_10}, find the first uncovered link of the just-visited node and mark it as covered. Then, select the other node connected to the link and add it to the path. This newly added node becomes the current node, and the process repeats until the current node has no uncovered connected links. Note that node 1 is chosen as the initial node.
\item {\textit{Restriction:}} When there are no uncovered connected links, return to the previously visited node that still has uncovered connected links. Create a new path and add the node to it. Then, repeat step 2. If the path length exceeds the threshold $l_{th}$, complete the path planning of the current path and add the node to the new path.
\item {\textit{Termination test:}} When the termination condition is satisfied, stop and complete all path planning tasks. The termination condition can also be when the network telemetry system covers the entire network topology or the maximum running time is achieved.
\end{enumerate}

\begin{algorithm}
\caption{Auxiliary Probes Path Deployment Algorithm}
\label{alg1}
\begin{algorithmic}[1]
\REQUIRE Graph $G$ and starting node $v_0$
\ENSURE A set of auxiliary probe paths
\STATE{\bf{Function:}}{~~}{APPD}{(node $v$ , bool $flag$)}
\IF{$flag$ }
\STATE Create a new path as the current path.
\STATE Add $v$ to the current path, $n \gets 1$.
\ENDIF
\STATE $flag \gets 0$.
\FORALL{$i \in V\left( v \right)$}
\IF{$ \left( v,i \right)\ne None $}
\STATE $\left( v,i \right)=\left( i,v \right)=None$.
\IF{$ flag $}
\STATE Create a new path as the current path. 
\STATE Add $v$ and $i$ to the current path, $n \gets 2$.
\ELSE 
\IF {$n<l_{th}$}
\STATE Add $i$ to the current path, $n \gets n+1$.
\ELSE  
\STATE $flag \gets 1$.
\ENDIF
\ENDIF
\STATE $flag \gets$APPD$(v,flag)$.
\ENDIF
\STATE {\bf return} $true$.
\ENDFOR
\end{algorithmic}
\end{algorithm}

\vspace{-0.5cm}
\section{Dynamic Probe Path Deployment System}
\label{section5}
In this section, we first analyze the problem of the dynamic probe path deployment. Then, we propose an end-to-end DRL framework and apply it to design the DPPD algorithm. Finally, a transfer learning method is employed to reduce the training time of the DRL model. 
\vspace{-0.5cm}

\subsection{Problem Analysis}

This subsection analyzes the problem of the dynamic probe path deployment based on section \ref{section4A}. 
For a specific query, we assume that the number of dynamic probe paths in the network telemetry system is $Q$. We denote the $q$-th path as ${{{d}}_{q}}=\left[ {{v}'_{q,1}},{{v}'_{q,2}},\cdots ,{{v}'_{q,N_{q}}} \right]$, where ${v}'_{q,i}$ denotes the $i$-th node that path $d_q$ passes through and $N'_{q}$ is the number of nodes in path $d_q$. The set of links that the $q$-th path passes through is denoted by ${L}'_{q}$.

The network application not only determines the types of metadata required for the query, but also puts forward requirements for which links should be probed. We refer to the subnet that supports the network application as a service network, which is composed of switches and links that support the network application. The set of links in the service network is denoted by $S$, $S\subseteq E$. DPs obtain the metadata required for the query from the switches they pass through.
To obtain comprehensive service network information for the controller, the telemetry targets of DPs must include the entire service network, which can be represented as

\begin{IEEEeqnarray}{rCl} %��ʽ5
\label{form_5}
{\!}
\begin{split}
S\subseteq \bigcup\limits_{q=1}^{Q}{{{L}'_{q}}}.\\
\end{split}
\end{IEEEeqnarray}

Furthermore, since DPs undertake the main task of network telemetry, the deployment of dynamic probe paths has a significant impact on the performance of the telemetry system.  During the path planning process, we need to meet users' various performance requirements as much as possible, including control overhead, latency, coverage, computing resources, storage resources, bandwidth utilization, etc. 
We represent all performance indicators of a telemetry task as the set $F=\left\{ {{f}_{1}},{{f}_{2}},\cdots,{{f}_{i}},\cdots,{{f}_{m}} \right\}$, where ${{f}_{i}}\in F$ represents the numerical value of a performance indicator.
Because the importance of each performance indicator is different, we set the weight of the performance indicator ${{f}_{i}}$ to $w_i$, $w_i \in W$, where $W=\left\{ {{w}_{1}},{{w}_{2}},\cdots,{{w}_{i}},\cdots,{{w}_{m}} \right\}$.
We add the products of each performance indicator and its weight as the optimization objective for dynamic probe path planning, which can be denoted as
\begin{IEEEeqnarray}{rCl} %��ʽ5
\label{form_5.5}
{\!}
\begin{split}
C=\sum\limits_{i=1}^{m}{{{w}_{i}}{{f}_{i}}},\\
\end{split}
\end{IEEEeqnarray}
where $C$ is the weighted telemetry revenue which is related to $F$ and $W$. Considering covering the entire service network, the dynamic probe path planning problem can be formulated as
\begin{subequations}%y优化问题
\label{form_6}
\begin{equation}
\tag{\ref{form_6}}
{~~}{\underset{{{d}_{q}},q=1,\cdots ,Q}{\mathop{\min }}\,\text{  }C=\sum\nolimits_{i=1}^{m}{{{w}_{i}}{{f}_{i}}}}
\end{equation}
\vspace{-1.8em}
\begin{align}
\mathrm{s.t.}{~~}
&{S\subseteq \bigcup\limits_{q=1}^{Q}{{{L}'_{q}}}}.\label{form_6.a}
\end{align}
\end{subequations}
We influence each performance indicator of the telemetry system by adjusting the paths of the DPs.
The constraint \eqref{form_6.a} ensures that the path of DPs can cover the entire service network, which means that the network information collected by the telemetry system can meet the requirements of the telemetry task.

Problem \eqref{form_6} represents a challenging multi-objective optimization problem, particularly in the context of multi-path planning for network telemetry systems. Given the complex and dynamic nature of modern networks, several factors make it difficult to solve this problem directly, including:

\begin{itemize}
\item Multiple optimization objectives: Network telemetry systems must balance multiple optimization objectives simultaneously. These objectives often conflict with one another, making it challenging to find optimal solutions that meet all requirements.
\item Diversity of telemetry tasks: Each telemetry task has unique requirements and constraints, requiring algorithms to be flexible to different optimization objectives and scenarios.
\item Dynamic network environment: Network topologies and traffic can change rapidly, requiring telemetry systems to adapt quickly to changing conditions while minimizing downtime and other performance issues.
\end{itemize}

To address these challenges, the proposed algorithm based on DRL offers several significant advantages over classical heuristic algorithms. 
Firstly, it automatically adjusts the deployment plan according to diverse users' requirements, making it more robust than traditional heuristic algorithms. Secondly, it can quickly solve the same type of combinatorial optimization problem and adapt to dynamic changes in the network environment.

\vspace{-1em}
\subsection{Deep Reinforcement Learning Model}

In this subsection, we introduce the DRL model \cite{26}, which is shown in Fig. \ref{fig5}.

\subsubsection{Model Structure Review}
\begin{figure}
\centering
\setlength{\abovecaptionskip}{0.cm}
\setlength{\belowcaptionskip}{-10 cm}
\includegraphics[width=8.5cm]{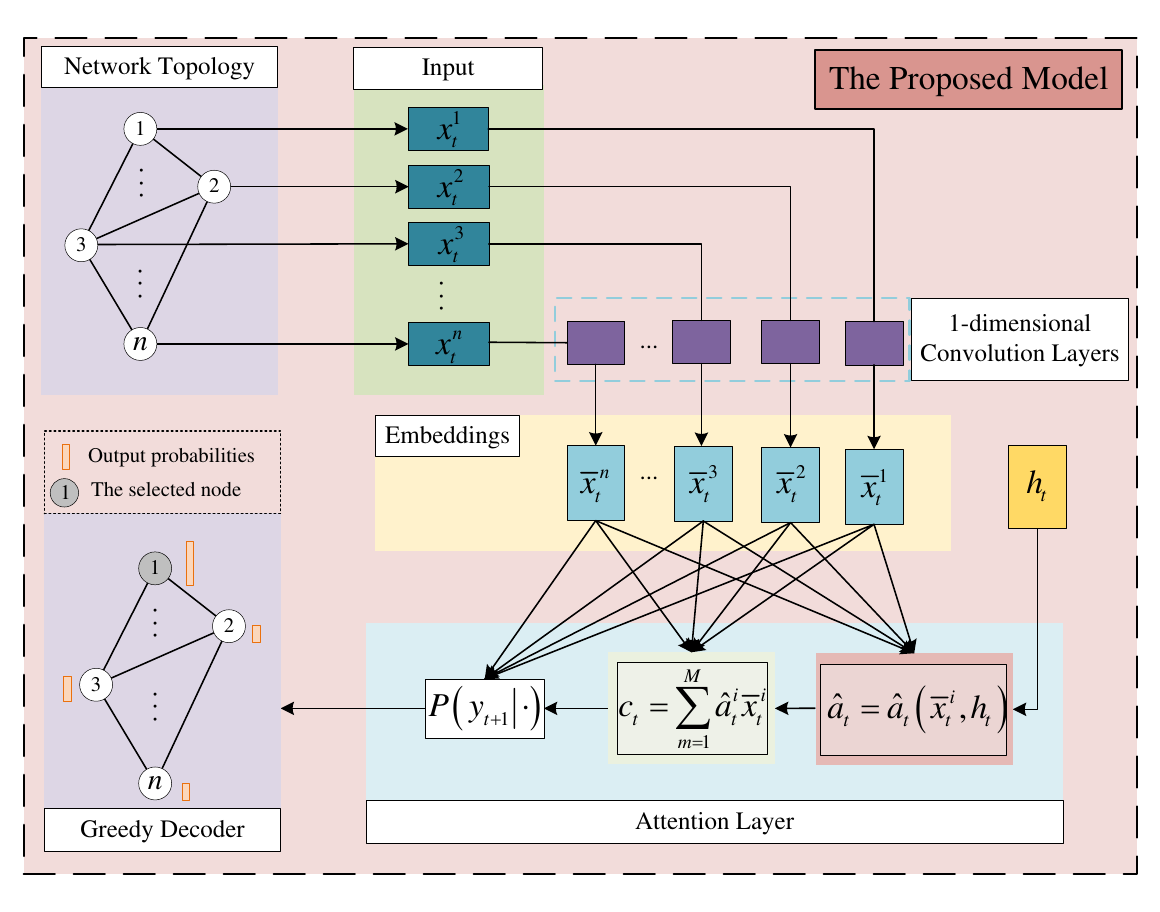}
\caption{The proposed model for the DPPD algorithm. At each time step $t$, the model contains an embedding layer and an attention layer.  }
\label{fig5}
\vspace{-0.1cm}
\end{figure}

Problem \eqref{form_6} is a combinatorial optimization problem, and we assume that the input set is $\mathcal{X}\doteq \left\{ {{x}^{i}},i=0,1,\cdots,n \right\}$. Each input ${x}^{i}$ is a sequence of network information tuples, containing the connection of the node $i$ with other nodes, the latency information of each port, and so on. 

We begin by selecting an arbitrary input value ${{y}_{0}}\in {\mathcal{X}_{0}}$. At each decoding step $t$, $t\in \left[ 0, T \right]$, we select ${y}_{t+1}$ from the set of available inputs $\mathcal{X}_{t}$ and continue until constraint \eqref{form_6.a} is met, which means that no more demand need to be satisfied. The sequence generated by this process can be denoted as $\mathcal{Y}=\left\{ {{y}_{t}},t=0,\cdots ,{T}' \right\}$, where ${T}'$ is the sequeue length. We use ${\mathcal{Y}_{t}}=\left\{ {{y}_{0}},\cdots ,{{y}_{t}} \right\}$ to denote the decoded sequence at time $t$. We aim to make the generated stochastic policy as close as possible to the optimal ones.
Similar to \cite{sutskever2014sequence}, we adopt the probability chain rule to break down the probability $P\left( \left. \mathcal Y \right|{\mathcal{X}_{0}} \right)$ of generating the sequence $\mathcal{Y}$, which are expressed as follows:

\begin{IEEEeqnarray}{rCl} %��11
\label{form_11}
{\!}
\begin{split}
P\left( \left. \mathcal Y \right|{\mathcal{X}_{0}} \right)=\prod\limits_{t=0}^{T}{P\left( \left. {{y}_{t+1}} \right|{\mathcal{Y}_{t}},{\mathcal{X}_{t}} \right)},\\
\end{split}
\end{IEEEeqnarray}
where $P\left( \left. {{y}_{t+1}} \right|{\mathcal{Y}_{t}},{\mathcal{X}_{t}} \right)$ is computed by the attention mechanism, which will be described in detail in the next subsection. We denote the affine function that outputs an input-sized vector as $\operatorname{g}$ and the state of the RNN decoder as $h_t$, which summarizes the information of previously decoded steps ${{y}_{0}},\cdots,{{y}_{t}}$. $P\left( \left. {{y}_{t+1}} \right|{\mathcal{Y}_{t}},{\mathcal{X}_{t}} \right)$ can be expressed as
\begin{IEEEeqnarray}{rCl} %��13
\label{form_13}
{\!}
\begin{split}
{P\left( \left. {{y}_{t+1}} \right|{\mathcal{Y}_{t}},{\mathcal{X}_{t}} \right)=\text{softmax}\left( \operatorname g\left( {{h}_{t}},{\mathcal{X}_{t}} \right) \right).}\\
\end{split}
\end{IEEEeqnarray}
In addition, according to \cite{sutskever2014sequence}, we can express the recursive update of the problem representation as 

\begin{IEEEeqnarray}{rCl} %��12
\label{form_12}
{\!}
\begin{split}
{{\mathcal{X}}_{t+1}}=\operatorname{f}\left( {{y}_{t+1}},{{\mathcal{X}}_{t}} \right),\\
\end{split}
\end{IEEEeqnarray}
where the role of the state transition function $\operatorname{f}$ is to recursively update the problem representation.

The RNN encoder's complexity is high as it attends to the order of the input set, which is crucial for tasks like text translation where word combination and position impact translation accuracy. However, Problem \eqref{form_6}, where the input is the information of a set of nodes, and we do not need to pay attention to their order. Therefore, we simply remove the RNN encoder and use input embedding directly to reduce model complexity.
Fig. \ref{fig5} shows the proposed model. We map the input into a vector space, which may have multiple embeddings corresponding to different input elements that are shared. We utilize 1-dimensional convolutional layers To perform the embedding.

\subsubsection{Attention Mechanism}

The attention layer in Fig. \ref{fig5} shows the proposed model's attention mechanism, which is a commonly used structure for processing input. Similar to \cite{vinyals2015pointer}, to extract relevant information from the inputs at decoder step $i$, 
we utilize a content-based attention mechanism with a glimpse. The variable-length alignment vector ${{\hat{a}}_{t}}$ is used to compute this mechanism. $\bar{x}_{t}^{i}$ represents the embedded input $x_{t}^{i}$. Furthermore, ${{h}_{t}}\in {{\mathbb{R}}^{D}}$ denotes the memory state of the RNN cell at decoding step $t$. The variable-length alignment vector $\hat{a}_t$ can be computed as
\begin{IEEEeqnarray}{rCl} %��13
\label{form_13}
{\!}
\begin{split}
{\hat{a}_{t}}={\hat{a}_{t}}\left( \bar{x}_{t}^{i},{{h}_{t}} \right)=\text{softmax}\left( {{u}_{t}} \right),\\
\end{split}
\end{IEEEeqnarray}
where
\begin{IEEEeqnarray}{rCl} %��14
\label{form_14}
{\!}
\begin{split}
u_{t}^{i}=v_{\hat{a}}^{T}\tanh \left( {{W}_{\hat{a}}}\left[ \bar{x}_{t}^{i};{{h}_{t}} \right] \right),\\
\end{split}
\end{IEEEeqnarray}
${\hat{a}}_{t}$ determines the relevance of each input data point for the upcoming decoding step $t$, and the symbol ``${;}$'' denotes the concatenation of two vectors. The variables $v_{\hat{a}}$ and ${W}_{\hat{a}}$ are trainable variables. 

To compute the conditional probabilities, we first calculate the context vector $c_t$ with the embedded inputs, defined as
\begin{IEEEeqnarray}{rCl} %��15
\label{form_15}
{\!}
\begin{split}
{{c}_{t}}=\sum\limits_{m=1}^{M}{\hat{a}_{t}^{i}\bar{x}_{t}^{i}},\\
\end{split}
\end{IEEEeqnarray}

Then, we normalize the values with the softmax function to obtain the conditional probability as follows:
\begin{IEEEeqnarray}{rCl} %��16
\label{form_16}
{\!}
\begin{split}
P\left( \left. {{y}_{t+1}} \right|{{\mathcal Y}_{t}},{{\mathcal X}_{t}} \right)=\text{softmax}\left( \tilde{u}_{t}^{i} \right),\\
\end{split}
\end{IEEEeqnarray}
where
\begin{IEEEeqnarray}{rCl} %��17
\label{form_17}
{\!}
\begin{split}
\tilde{u}_{t}^{i}=v_{c}^{T}\tanh \left( {{W}_{c}}\left[ \bar{x}_{t}^{i};{{c}_{t}} \right] \right),\\
\end{split}
\end{IEEEeqnarray}
and the variables $v_{c}$ and ${W}_{c}$ are also trainable. 

\subsubsection{Training Method}
\begin{algorithm}
\caption{Reinforcement Learning Algorithm}
\label{alg2}
\begin{algorithmic}[1] 
\STATE \textbf{Initialization:} Initialize the actor network and critic network with random weights ${{\theta }}$ and ${{\delta }}$. 
\FOR {$i=1,2,\cdots ,epoch$}
\STATE Reset gradients. $d\theta \leftarrow 0$, $d\delta \leftarrow 0$
\STATE Sample instances  from set $\mathbf{M}$.
\FORALL {instances $\mathbf{m}=1,2,\cdots ,batch$}
\STATE $t\leftarrow0$.
\WHILE{termination condition is not reached,}
\STATE Choose the next node according to the output probabilities $P\left( \left. {{y}_{t+1}} \right|{{\mathcal Y}_{t}},{{\mathcal X}_{t}} \right)$.
\STATE Get the new state ${\mathcal X}_{t+1}$ .
\STATE $t\leftarrow t+1$.
\ENDWHILE
\STATE Compute the reward ${{R}^{\mathbf{m}}}$ based on the generated policy.
\ENDFOR
\STATE Compute $d\theta$ and $d\delta$ according to the rewards.
\STATE $d\theta \leftarrow \frac{1}{batch}\sum\limits_{\mathbf{m}=1}^{batch}{\left( {{R}^{\mathbf{m}}}-V\left( X_{0}^{\mathbf{m}};\delta  \right) \right)}{{\nabla }_{\theta }}\log P\left( \left. {{Y}^{\mathbf{m}}} \right|X_{0}^{\mathbf{m}} \right)$
\STATE $d\delta \leftarrow \frac{1}{batch}{{\sum\limits_{\mathbf{m}=1}^{batch}{{{\nabla }_{\delta }}\left( {{R}^{\mathbf{m}}}-V\left( X_{0}^{\mathbf{m}};\delta  \right) \right)}}^{2}}$
\STATE Update $\theta$ and $\delta$ according to $d\theta$ and $d\delta$.
\ENDFOR

\end{algorithmic}
\end{algorithm}

We use the classical policy gradient approach to train the network. These methods use the estimated gradient of the expected return concerning the policy parameters to iteratively improve the policy, which is a standard training method in reinforcement learning.
The policy gradient algorithm consists of two networks: (i) an actor network that predicts the probability distribution over the next action at any given decision step; 
(ii) a critic network that estimates the reward for any problem instance from a given state.
The critic network consists of a dense layer with ReLU activation and a linear layer with a single output.
In the Actor-Critic network, the critic network computes a weighted sum of the embedded inputs with the output probabilities of the actor network. 
To optimize the selected action in the current state, the actor network updates its network parameters through backpropagation. 

Similar to \cite{bello2017neural}, the details of our training method are shown in Algorithm \ref{alg2}. 
We consider a class of problems whose set of instances is represented by $\mathbf{M}$. Random weights ${{\theta }}$ and ${{\delta }}$ are respectively used to initialize the actor network and critic network. 
At the beginning of the training, we first obtain the problem instances from the set $\mathbf{M}$, where the variable $batch$ represents the number of instances. For each instance $\mathbf{m}$, we generate feasible sequences based on the output probabilities as the current policy. 
After the termination condition is reached, we use the generated policy to compute the reward ${{R}^{\mathbf{m}}}$. Then, we compute the policy gradient to update the actor network and critic network, where $V\left( X_{0}^{\mathbf{m}};\delta  \right)$ is the reward approximation calculated by the critic network. In general, the reinforcement learning algorithm provides an appropriate paradigm for training neural networks for combinatorial optimization problems with similar structures.

\vspace{-1.em}
\subsection{Dynamic Probe Path Deployment (DPPD) Algorithm}
In this subsection, we design the DPPD algorithm, a key component of AdapINT that is designed based on our proposed DRL model.  

We first create instances for training. To make our model adaptable to a variety of networks, we assume that whether a switch is directly connected to another switch in the network is completely random. The content of the input $\mathcal{X}$ can be adjusted to suit telemetry requirements.
In this paper, for convenience, ${x}^{i}\in \mathcal{X}$ contains the connection status of the node $i$ to other nodes and the latency information of each port.
\subsubsection{State}The state of the DRL network $s_t$ at the end of time slot $t$ is defined as
\begin{IEEEeqnarray}{rCl} %��18
\label{form_18}
{\!}
\begin{split}
{{s}_{t}}=\left\{ {{{\tilde{n}}}_{t}},{{E}_{t}} \right\},{{\tilde{n}}_{t}}\in V\bigcup \left\{ 0 \right\},t=0,\cdots ,T,\\
\end{split}
\end{IEEEeqnarray}
where $E_t$ represents the set of links that still have telemetry requirements at time slot $t$. ${\tilde{n}}_{t}\in V$ represents the current node index. ${\tilde{n}}_{t}=0$ represents that the algorithm is in the state of creating a new path.

\subsubsection{Action}At state $s_t$, our actions include adding a node to the path and creating a new path.
To represent the action of creating a new path, we introduce a virtual variable ``$0$". When the decoder selects the virtual variable ``$0$" as the next action, it means that we should create a new dynamic probe path. Thus, the actions at state $s_t$ can be denoted as

\begin{IEEEeqnarray}{rCl} %��19
\label{form_19}
{\!}
\begin{split}
{{a}_{t}}=\left\{ \left. i \right|\exists \left( {{{\tilde{n}}}_{t}},i \right)\in {{E}_{t}} \right\}\bigcup \left\{ 0 \right\},t=0,\cdots ,T.
\end{split}
\end{IEEEeqnarray}
As shown in Fig. \ref{fig5}, we use a greedy decoder to select actions, which can effectively improve the quality of the solution. 
Therefore, the action with the highest probability is selected at each decoding step. 
Then, the element ${\tilde{n}}_{t+1}$ is defined as the action with the highest probability of the current time slot, and $E_{t+1}$ also needs to update the state based on this action.

Moreover, since too many actions can be selected, we need to mask infeasible actions to train faster. We use a masking scheme to force some infeasible solutions to be masked and set their log probability to $-\infty $. Specifically, we enforce masking on the following nodes: $(i)$ nodes not connected to the current node. $(ii)$ nodes that have completed their telemetry requirements. Indeed, using a masking scheme can significantly reduce the solution space, resulting in faster attainment of better solutions. In addition to improving training efficiency, the masking scheme enhances solution reliability. However, this scheme may also result in the exclusion of optimal solutions. Therefore, it is essential to strike a balance between achieving optimality and ensuring reliability.
\subsubsection{Reward}We set the reward function of the model according to Problem \eqref{form_6}, which can be expressed as
\begin{IEEEeqnarray}{rCl} %��20
\label{form_20}
{\!}
\begin{split}
r=\underset{{{d}_{q}}}{\mathop{\min }}\,\sum\nolimits_{i=1}^{m}{{{w}_{i}}{{f}_{i}}},q=1,\cdots ,Q.\\
\end{split}
\end{IEEEeqnarray}
When users' requirements change, users only need to modify the reward function according to their requirements. They can retrain a model capable of meeting new requirements without any manual calculation.

\vspace{-0.5cm}
\subsection{Transfer Learning for DPPD Algorithm}
\begin{figure}
\centering
\setlength{\abovecaptionskip}{0.cm}
\setlength{\belowcaptionskip}{-10 cm}
\includegraphics[width=8.5cm]{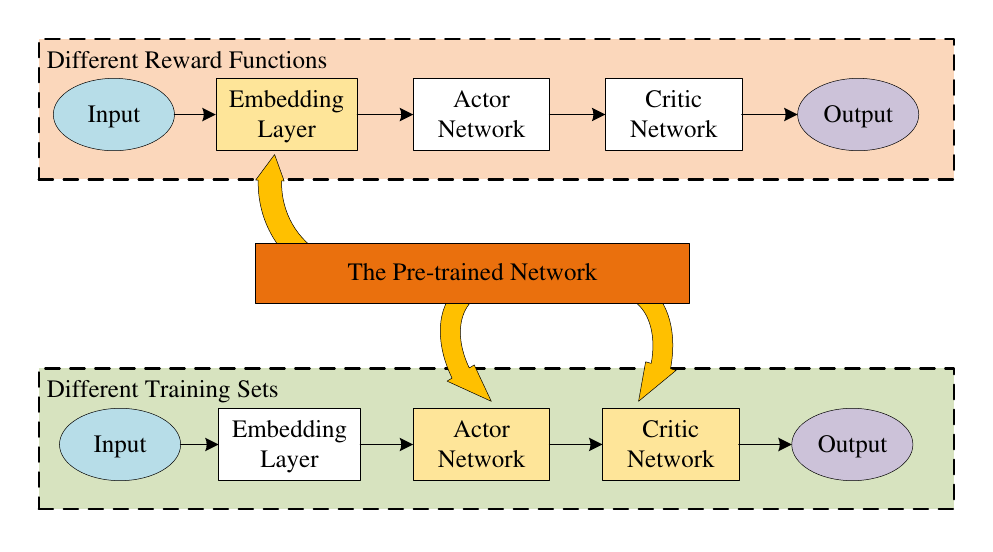}
\caption{Transfer learning model. }
\label{figTL}
\vspace{-0.1cm}
\end{figure}

When telemetry requirements change, the DPPD algorithm must retrain the DRL model using a new reward function, which incurs additional costs. Furthermore, if the network environment significantly varies, such as in terms of the network scale, creating multiple training sets to train the network also results in extra expenses. To address these issues, we implemented a transfer learning approach \cite{TL} to assist in training the model. This allowed the DPPD algorithm to solve different problems more quickly and be applied to various network environments.

Fig. \ref{figTL} presents the transfer learning process, which deals with different reward functions and training sets. Algorithm \ref{alg3} details the pseudocode for the transfer learning approach, which shares a similar training method shown in Fig. \ref{fig5}. To accommodate various reward functions, we need to load the pre-trained parameters of the embedding layer and retrain the actor and critic networks. Similarly, when facing different training sets, we should load the pre-trained parameters of the actor and critic networks and retrain the embedding layer. In cases where both the reward function and training set are different, we can utilize the transfer learning approach to retrain the entire model. This is because even though all networks require retraining, some parameters from specific networks can still be reused in the new model, thus reducing the training time. Once the entire network converges, we end the model training.

\begin{algorithm}[H]
\caption{Transfer Learning Training Algorithm}
\label{alg3}
\begin{algorithmic}[1] 
\renewcommand{\algorithmicrequire}{ \textbf{Initialization:} Load the pre-trained model parameters.}
\REQUIRE        % 实际显示Input 
\renewcommand{\algorithmicrequire}{ \textbf{Function:} Training for different reward functions}
\REQUIRE        % 实际显示Input
\STATE \textbf{Input:} The training set with the different reward functions.
\STATE Get the model parameters from the pre-trained model.
\STATE Retrain the actor network and the critic network.
\WHILE{Termination condition is not met}
\STATE Train the deep reinforcement model.
\ENDWHILE
\STATE \textbf{Output:} The re-trained network.
\renewcommand{\algorithmicrequire}{ \textbf{Function:} Training for different training sets}
\REQUIRE        % 实际显示Input
\STATE \textbf{Input:} The different training set.
\STATE Get the model parameters from the pre-trained model.
\STATE Retrain the embedding layers.
\WHILE{Termination condition is not met}
\STATE Train the deep reinforcement model.
\ENDWHILE
\STATE \textbf{Output:} The re-trained network.
\end{algorithmic}
\end{algorithm}

\vspace{-1.5em}
\section{Performance Evaluation}
\label{section6}

In this section, we simulate the AdapINT in Python 3 on the platform with an Intel (R) Core (TM) i7-7700k CPU @ 4.20ghz 4.20GHz machine equipped with 8GB RAM. 
The evaluation of AdapINT consists of two parts: evaluating the APPD algorithm and the DPPD algorithm.

For the APPD algorithm, we evaluated the coverage of the algorithm and simulated a fault localization scenario on auxiliary probe paths. For the proposed DPPD algorithm, we compared the telemetry performance of our proposed DPPD algorithm with three traditional algorithms, namely the DFS algorithm, Euler Trail/Circuit, and Latency-Constrained algorithm \cite{12}, under various telemetry requirements. Next, we briefly introduce various traditional algorithms. 
\begin{itemize}
\item DFS algorithm is a graph traversal technique that starts at an arbitrary node and explores as far as possible along each branch before backtracking. This algorithm can be implemented using a stack or recursion to keep track of the nodes to visit and the order in which to visit them. The advantage of the DFS algorithm is that it has extremely low time complexity. 
\item The Euler Trail/Circuit algorithm is a graph theory algorithm whose purpose is to make the number of paths reach a theoretical minimum. Specifically, each path extracted from the graph should begin at an odd vertex and end at another odd vertex. Removing one such path from the graph will eliminate a pair of odd vertices. Euler Trail/Circuit is based on the idea of extracting paths iteratively between pairs of odd vertices until all edges or vertices are removed from the original graph. This algorithm has the advantage of minimizing the number of generated paths, which makes it an efficient option for reducing telemetry overhead.
\item 
The Latency-Constrained algorithm controls telemetry latency by setting limits on the maximum latency threshold $T_{\max}$ during path planning, such as IntOpt and NetView \cite{13, 14}. Since the setting of the maximum latency threshold $T_{\max}$ is subjective, for convenience, we set the threshold as a linear function related to the number of switches, and apply it to the DFS algorithm.

\end{itemize}

We analyze the performance of AdapINT on random topologies without isolated nodes. Considering the clogged drain effect and the straggling herd effect \cite{27}, we refer to the prototype on Mininet and adopt the latency setting scheme proposed in \cite{el2021evaluating}.
It is difficult for traditional algorithms to support complex network topologies with multiple services. To simplify the evaluation, we assumed that all switches in the network provided the same service and determined the metrics that users care about according to different scenarios.
These metrics were not assigned specific weights since their relative importance may vary among different scenarios.
\begin{figure}
\centering
\setlength{\abovecaptionskip}{0.cm}
\setlength{\belowcaptionskip}{-10 cm}
\includegraphics[width=6.5cm]{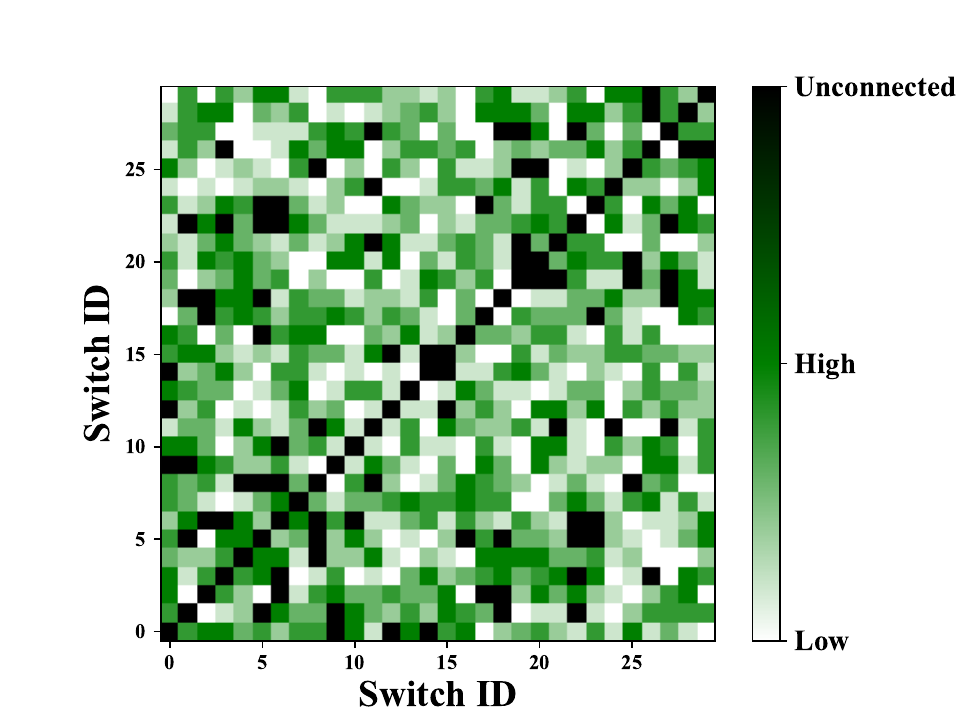}
\caption{Visibility of basic network information. }
\label{fig7}
\vspace{-0.1cm}
\end{figure}

\begin{figure}
\centering
\setlength{\abovecaptionskip}{0.cm}
\setlength{\belowcaptionskip}{-10 cm}
\includegraphics[width=6.5cm]{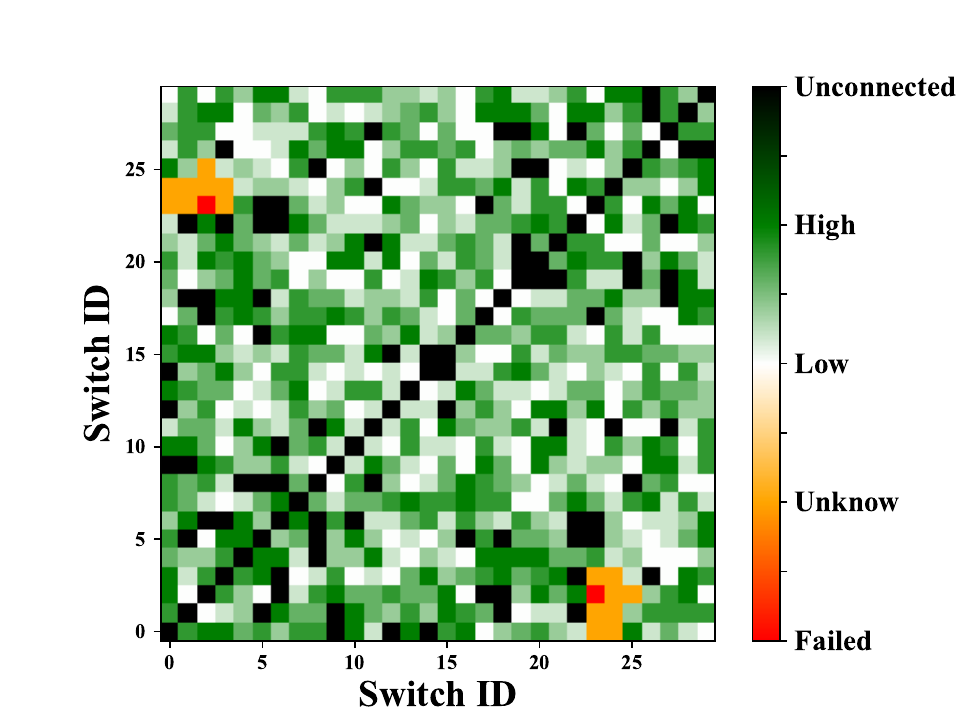}
\caption{Network fault location information. }
\label{fig8}
\vspace{-0.1cm}
\end{figure}
\vspace{-1em}
\subsection{Evaluation of APPD Algorithm}
In this subsection, we aim to demonstrate the reliability of our APPD algorithm. To achieve this, we simulate a network consisting of 30 switches. The basic information collected by the auxiliary probe is the link latency, which is vital for assessing the performance of our algorithm. By analyzing the results obtained from this experiment, we can determine whether the algorithm can deploy auxiliary probe paths efficiently in a real-world network.

Fig. \ref{fig7} is a grid diagram composed of basic network information collected by APs, which can intuitively visualize network-wide traffic loads. The black grid indicates no link connection, while the green grid indicates the load according to the depth of the colour. Through APs, we can efficiently gather network-wide critical information. As demonstrated in Fig. \ref{fig7}, we can quickly identify areas experiencing heavy congestion through the darker colouration.
Fig. \ref{fig8} shows the network information grid diagram for random link failures. The red grids signify faulty links, while the yellow grids represent unknown links. We can observe a significant reduction in the range of faults that the controller needs to troubleshoot, thereby enhancing the efficiency of fault diagnosis. These results demonstrate the effectiveness of our proposed APPD algorithm, which is instrumental in facilitating telemetry analysis.

\vspace{-0.3em}
\subsection{Evaluation of DPPD Algorithm}
In this subsection, we analyze the performance of the DPPD algorithm in various scenarios. During training, we created 64000 instances of the network topology whose number of nodes is 30. Our model was trained for 50 epochs with a batch size of 1280. Both the Actor network and the critic network have a learning rate of 0.0001.
\begin{figure*} %这里使用的是强制位置，除非真的放不下，不然就是写在哪里图就放在哪里，不会乱动
	\centering
	\begin{minipage}{0.32\linewidth}
		\centering
		\includegraphics[width=6cm]{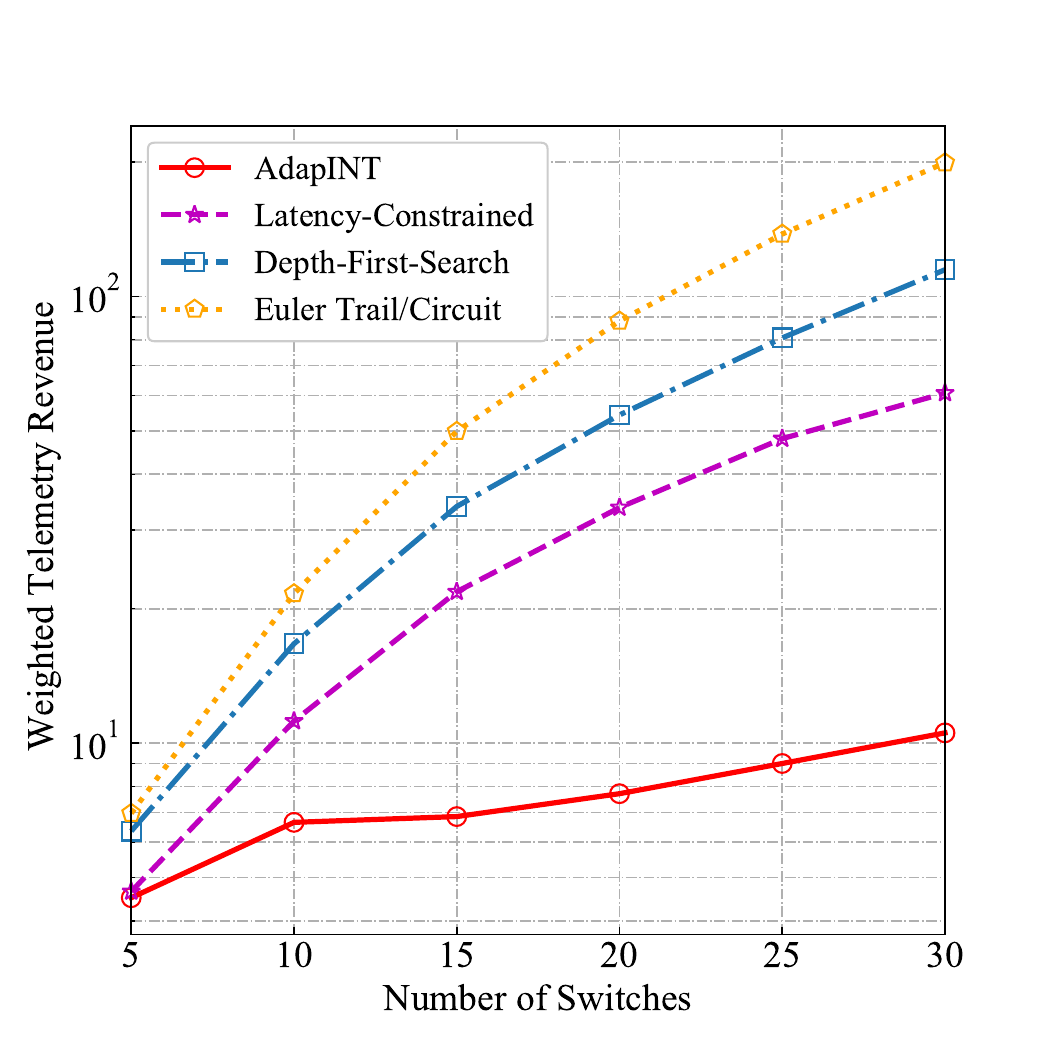}
		\caption{The weighted telemetry revenue in LO-L telemetry task. }
		\label{fig10}
	\end{minipage}
	%\qquad
	\begin{minipage}{0.32\linewidth}
		\centering
		\includegraphics[width=6cm]{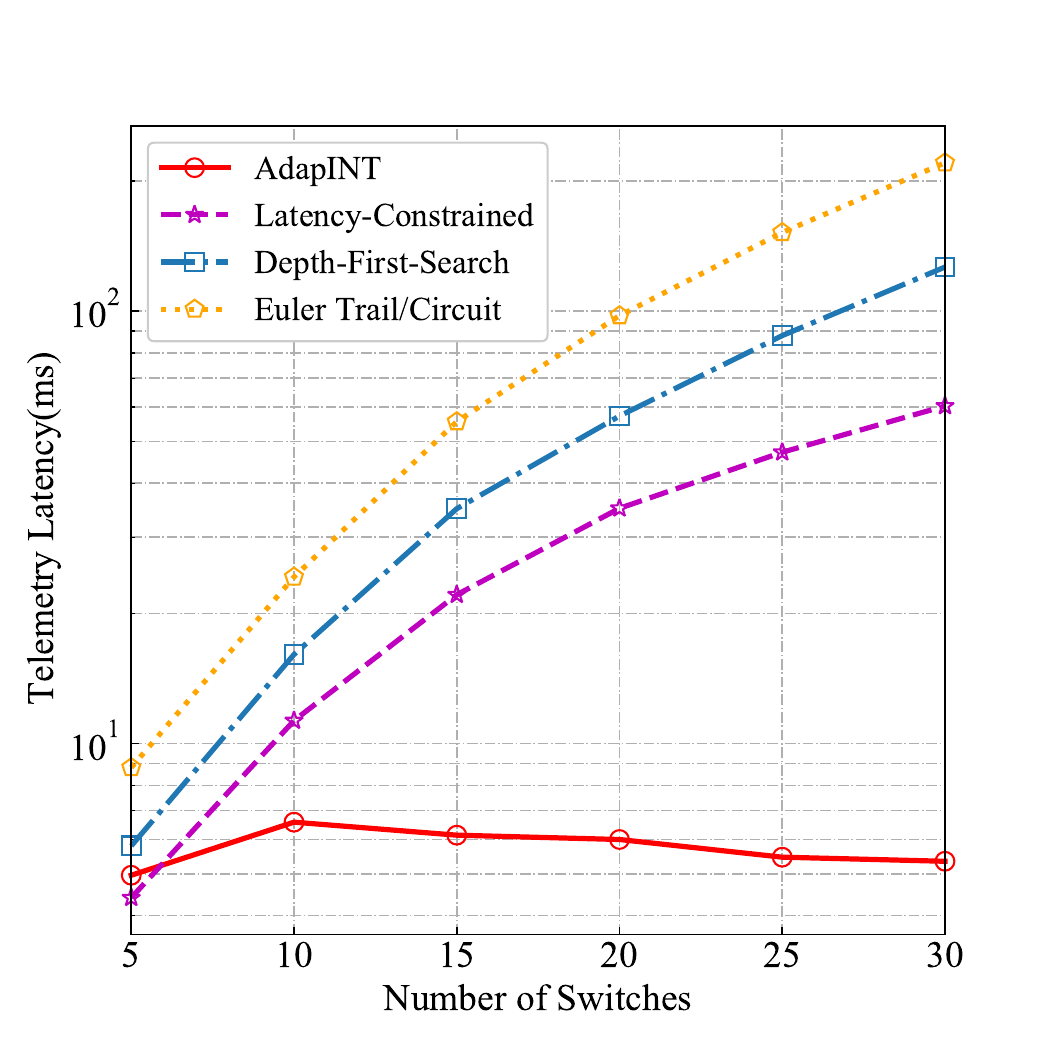}
		\caption{Telemetry latency in LO-L telemetry task.  }
		\label{fig11}
	\end{minipage}
	%\qquad
	\begin{minipage}{0.32\linewidth}
		\centering
		\includegraphics[width=6cm]{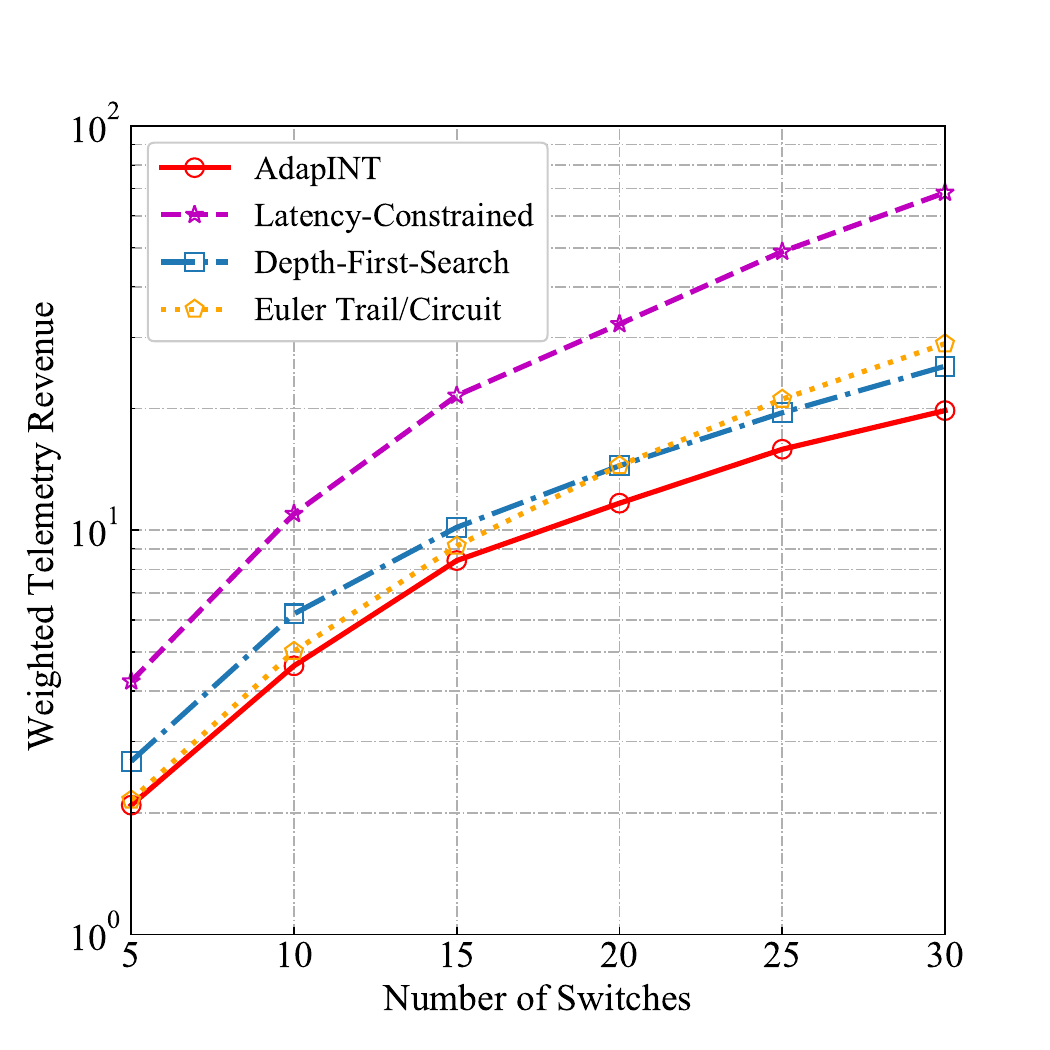}
		\caption{The weighted telemetry revenue in LO-O telemetry task. }
		\label{fig12}
	\end{minipage}
\end{figure*}

\begin{figure*} %这里使用的是强制位置，除非真的放不下，不然就是写在哪里图就放在哪里，不会乱动
	\centering
	\begin{minipage}{0.32\linewidth}
		\centering
		\includegraphics[width=6cm]{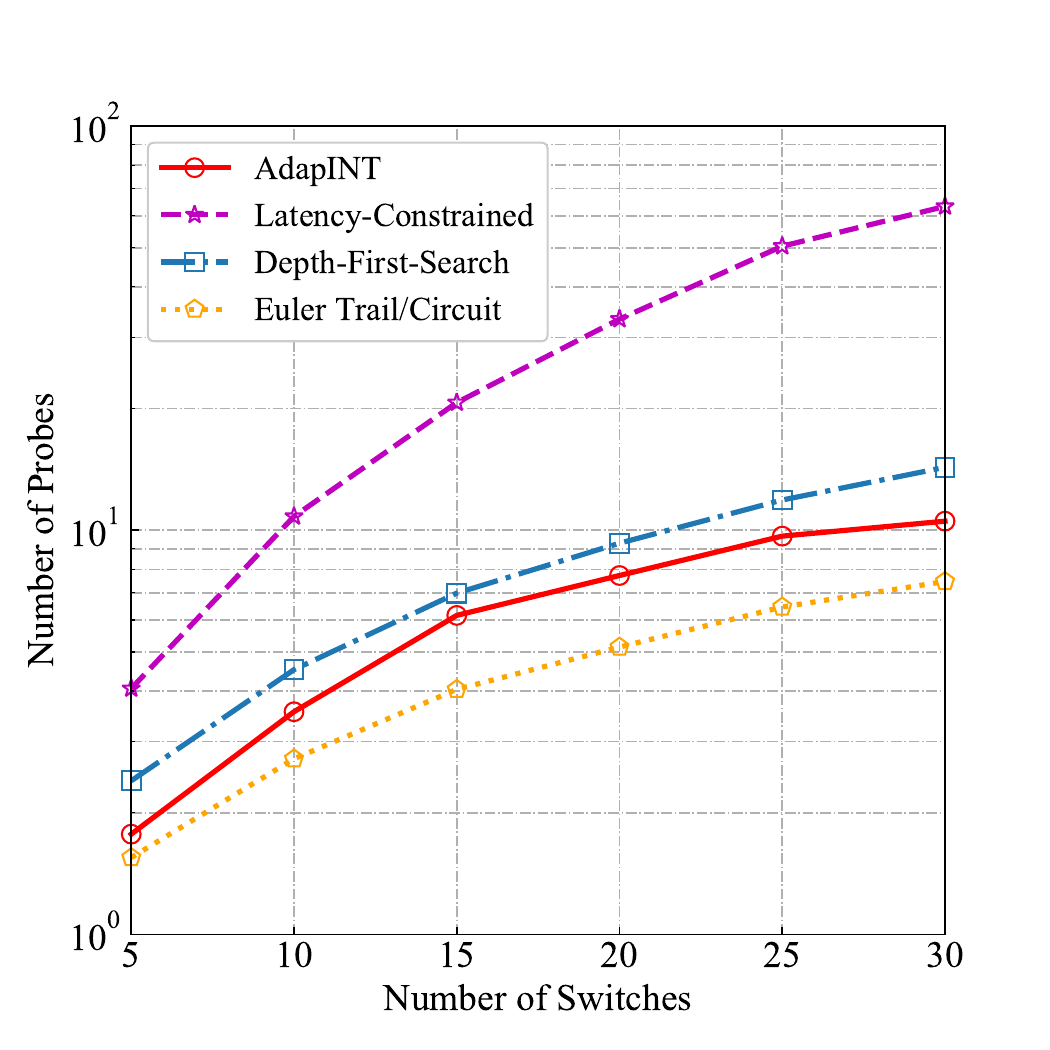}
		\caption{Telemetry overhead in LO-O telemetry task. }
		\label{fig13}
		\vspace{-0.1cm}
	\end{minipage}
	%\qquad
	\begin{minipage}{0.32\linewidth}
		\centering
		\includegraphics[width=6cm]{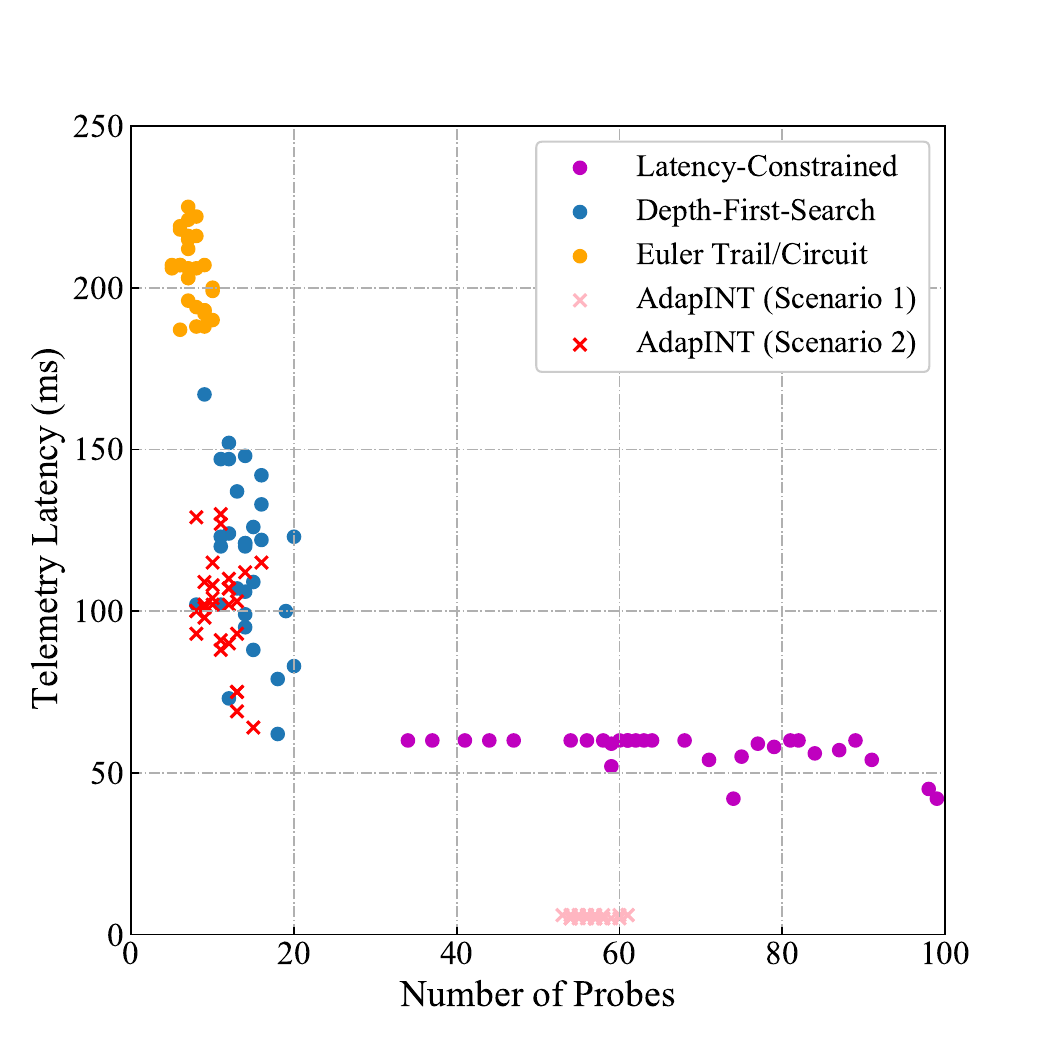}
		\caption{Telemetry latency and the number of probes in different algorithms. }
		\label{fig14}
	\end{minipage}
	%\qquad
	\begin{minipage}{0.32\linewidth}
		\centering
		\includegraphics[width=6cm]{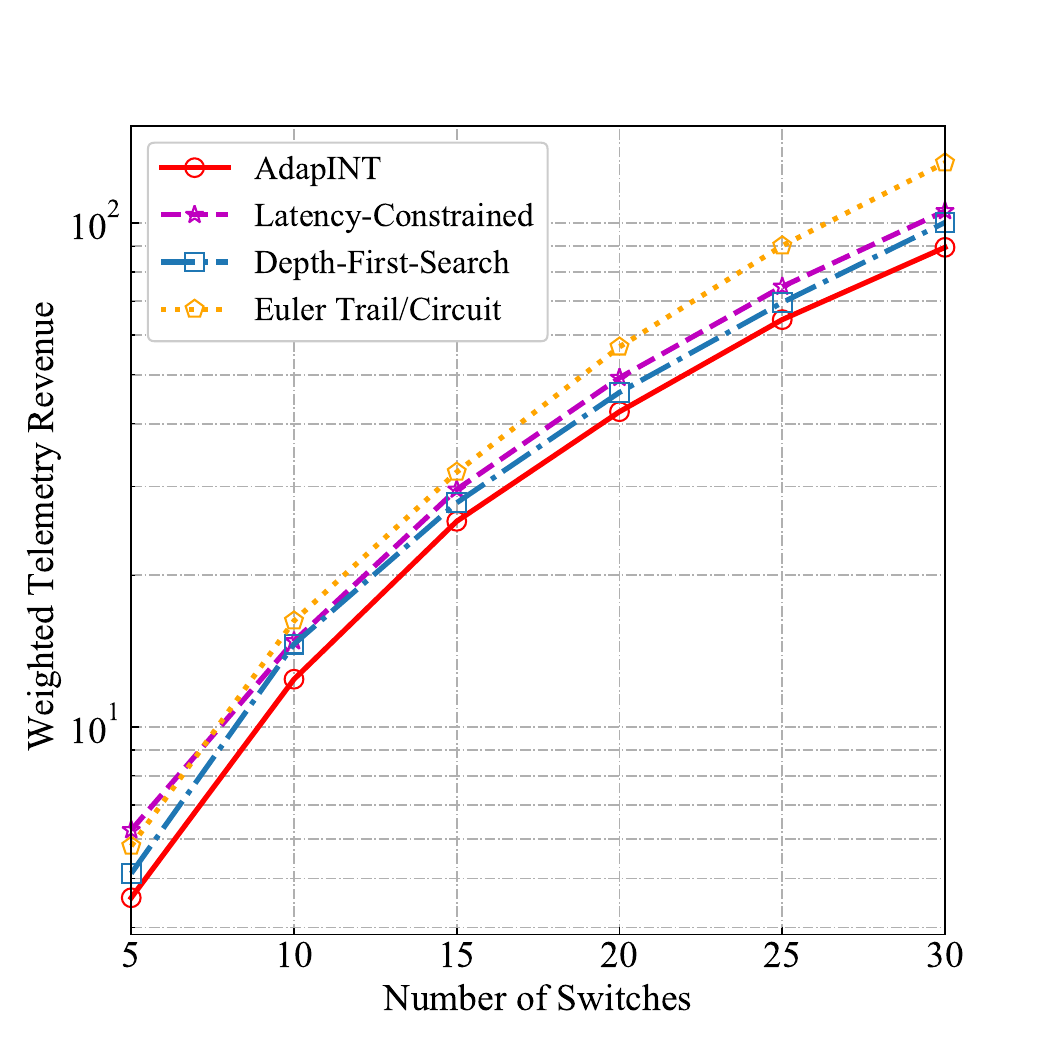}
		\caption{The weighted telemetry revenue in LOB-LO telemetry task.}
		\label{fig9}
		\vspace{-0.1cm}
	\end{minipage}

\end{figure*}

To demonstrate that AdapINT can meet various telemetry requirements, we evaluated the network telemetry performance in the following three scenarios:

$(i)$ Scenario 1. Consider scenarios such as video conferencing and online gaming. Weighted telemetry revenue comprises control overhead and telemetry latency, and users are more concerned about telemetry latency. We refer to telemetry tasks in this scenario as Latency and Control Overhead-Prioritized Latency-Aware telemetry tasks (LO-L telemetry tasks).

$(ii)$ Scenario 2. Consider scenarios such as cloud computing services. Weighted telemetry revenue is also composed of control overhead and telemetry latency, but users pay more attention to control overhead. We refer to telemetry tasks in this scenario as Latency and Control Overhead-Prioritized Control Overhead-Aware telemetry tasks (LO-O telemetry tasks).

$(iii)$ Scenario 3. Consider scenarios such as intelligent transportation systems and the Internet of Things. Weighted telemetry revenue consists of three metrics: control overhead, telemetry latency, and bandwidth utilization. Users are more concerned with control overhead and telemetry latency. We refer to telemetry tasks in this scenario as Latency, Control Overhead, and Bandwidth Utilization-Prioritized Latency and Control Overhead-Aware telemetry tasks (LOB-LO telemetry tasks).
Next, we analyze the performance of AdapINT in three scenarios to illustrate the advantages of the DPPD algorithm in various optimization objectives.

For convenience, we denote the control overhead weight as $w_1$, the latency weight as $w_2$, and the bandwidth utilization weight as $w_3$. 
To evaluate control overhead, we focused on the controller and identified probe collection as the main cause of such overhead. Specifically, as the number of probes that need to be processed per unit of time increases, telemetry overhead also increases. Therefore, we estimated telemetry overhead by measuring the number of probes that need to be processed.
As for telemetry latency, we considered the time it takes for the controller to collect all available probes before performing telemetry analysis. In particular, we determined the telemetry latency of AdapINT by identifying the return time of the probe from the longest path.
Then, bandwidth utilization is related to the total amount of network information collected by all probes.

\subsubsection{Scenario 1}Since telemetry latency is important, we set the latency weight to 0.9, $w_2=0.9$, and the overhead weight to 0.1, $w_1=0.1$. Fig. \ref{fig10} shows the weighted telemetry revenue of various algorithms in the LO-L telemetry tasks.
The weighted telemetry revenue of all algorithms increases with the size of the network topology, as larger topologies require more probes and longer paths. Additional probes lead to heightened network overhead, while longer paths result in longer telemetry latency. We observe that AdapINT can achieve the weighted telemetry revenue at least 30$\%$ lower than traditional algorithms. Furthermore, due to the telemetry latency constraints, the weighted telemetry revenue of the Latency-Constrained algorithm has been shown to perform exceptionally well, second only to AdapINT. However, Euler Trail/Circuit, which uses the fewest number of probes and tends to generate ultra-long paths, may not be suitable for LO-L telemetry tasks. As such, there may be better choices than Euler Trail/Circuit in Scenario 1. Fig. \ref{fig11} shows the telemetry latency of various algorithms in detail. 
We can observe that even in large-scale networks, the telemetry latency of Dynamic INT remains low. Although there are fluctuations in telemetry latency, it is still the best among the four algorithms.
Although the latency-constrained algorithm limits the telemetry latency, the fixed constraint setting by subjective judgment has defects in adapting to different network topologies. Determining the optimal constraint for different network scenarios remains a challenging task.

\subsubsection{Scenario 2} We set the overhead weight to 0.9, $w_1=0.9$, and the latency weight to 0.1, $w_2=0.1$ in LO-O telemetry tasks. In Fig. \ref{fig12}, we show the weighted telemetry revenue of different algorithms. Similarly, AdapINT is also the best after retraining the model according to telemetry requirements. We can find that Euler Trail/Circuit and Depth-First Search excel in various network sizes, showcasing their strengths. Fig. \ref{fig13} shows the relationship between the number of probes and the size of the topologies. While Euler Trail/Circuit generates minor probes, its weighted telemetry revenue is not always the best and can sometimes be worse than the DFS algorithm. This is because Euler Trail/Circuit only considers the control overhead and does not consider telemetry latency. Although telemetry latency may not be critical in LO-O telemetry tasks, the significant telemetry latency associated with Euler Trail/Circuit ultimately affects its weighted telemetry revenue. Thus, compared to traditional algorithms, AdapINT achieves a better balance of multiple objectives.
Fig. \ref{fig14} displays a scatterplot of telemetry latency and the number of probes for various algorithms. It can be observed that the Latency-Constrained algorithm effectively controls telemetry latency within the threshold, while Euler Trail/Circuit can minimize the number of probes. Additionally, AdapINT can update path planning according to telemetry requirements by retraining the model, which is impossible with traditional algorithms. This demonstrates that AdapINT can solve multi-objective optimization problems with diverse telemetry requirements.
\begin{figure}
\centering
\setlength{\abovecaptionskip}{0.cm}
\setlength{\belowcaptionskip}{-10 cm}
		\includegraphics[width=6cm]{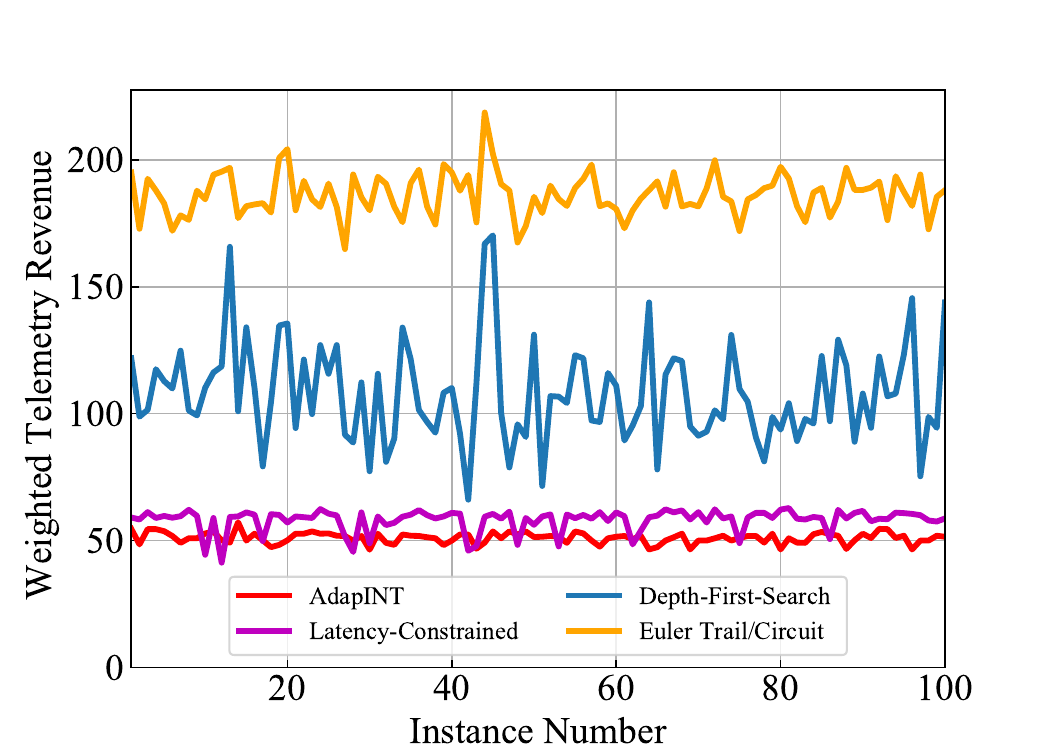}
		\caption{The weighted telemetry revenue in Scenario 1. }
		\label{fig15.1}
		\vspace{-0.1cm}
\end{figure}

\begin{figure}
\centering
\setlength{\abovecaptionskip}{0.cm}
\setlength{\belowcaptionskip}{-10 cm}
		\includegraphics[width=6cm]{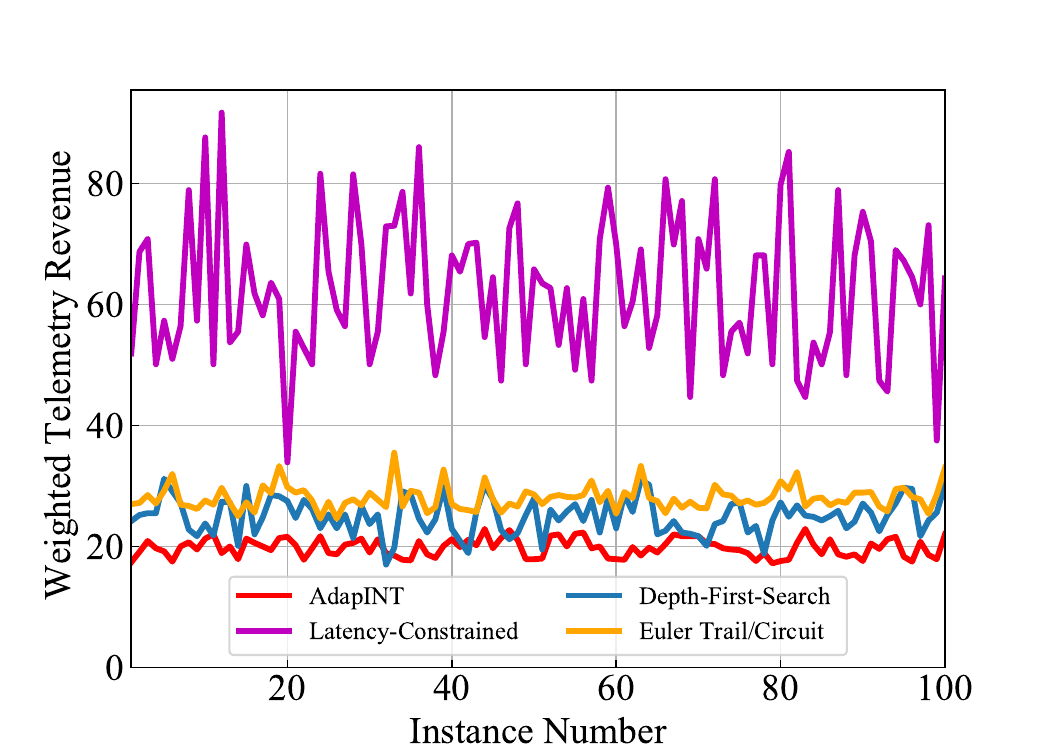}	
		\caption{The weighted telemetry revenue in Scenario 2. }	
		\label{fig15.2}
\end{figure}

\vspace{-0.cm}
\subsubsection{Scenario 3} Different from the above scenario, weighted telemetry revenue is composed of control overhead, telemetry latency and bandwidth utilization in LOB-LO telemetry tasks. Since both control overhead and telemetry latency are significant, we set the overhead weight and latency weight to 0.4, $w_1=0.4$ and $w_2=0.4$. Then, we set the bandwidth utilization weight to 0.2, $w_3=0.2$. Fig. \ref{fig9} shows the weighted telemetry revenue of various algorithms in LOB-LO telemetry tasks. We can find that AdapINT is still the best. Compared with the traditional algorithm, it can be seen that AdapINT can meet the user's requirements no matter whether the type or weight of the metric is changed. AdapINT is more suitable for multi-user or network environments where telemetry requirements change frequently.
Comprehensively analyzing the above three scenarios, we can find that AdapINT can meet users' requirements in various scenarios.

Moreover, AdapINT also has the adaptability to dynamically update policies based on the current environment and experience. Taking Scenario 1 and Scenario 2 as examples, we make the network topology and link latency change frequently to verify its adaptability.
Fig. \ref{fig15.1} shows the weighted telemetry revenue in Scenario 1.
Compared with other algorithms, we can find that AdapINT's weighted telemetry revenue is the most stable. The telemetry performance changes caused by dynamic network environments are relatively small.
As shown in Fig. \ref{fig15.2}, we can also reach the same conclusion in Scenario 2.
It is worth noting that AdapINT has the smallest variance in both scenarios, indicating that our proposed DPPD algorithm can adapt to changes in the network environment. This is attributed to the robustness of the DRL algorithm, which enables it to maintain outstanding performance even in the presence of noise and interference.

\begin{figure}
\centering
\setlength{\abovecaptionskip}{0.cm}
\setlength{\belowcaptionskip}{-10 cm}
		\includegraphics[width=6cm]{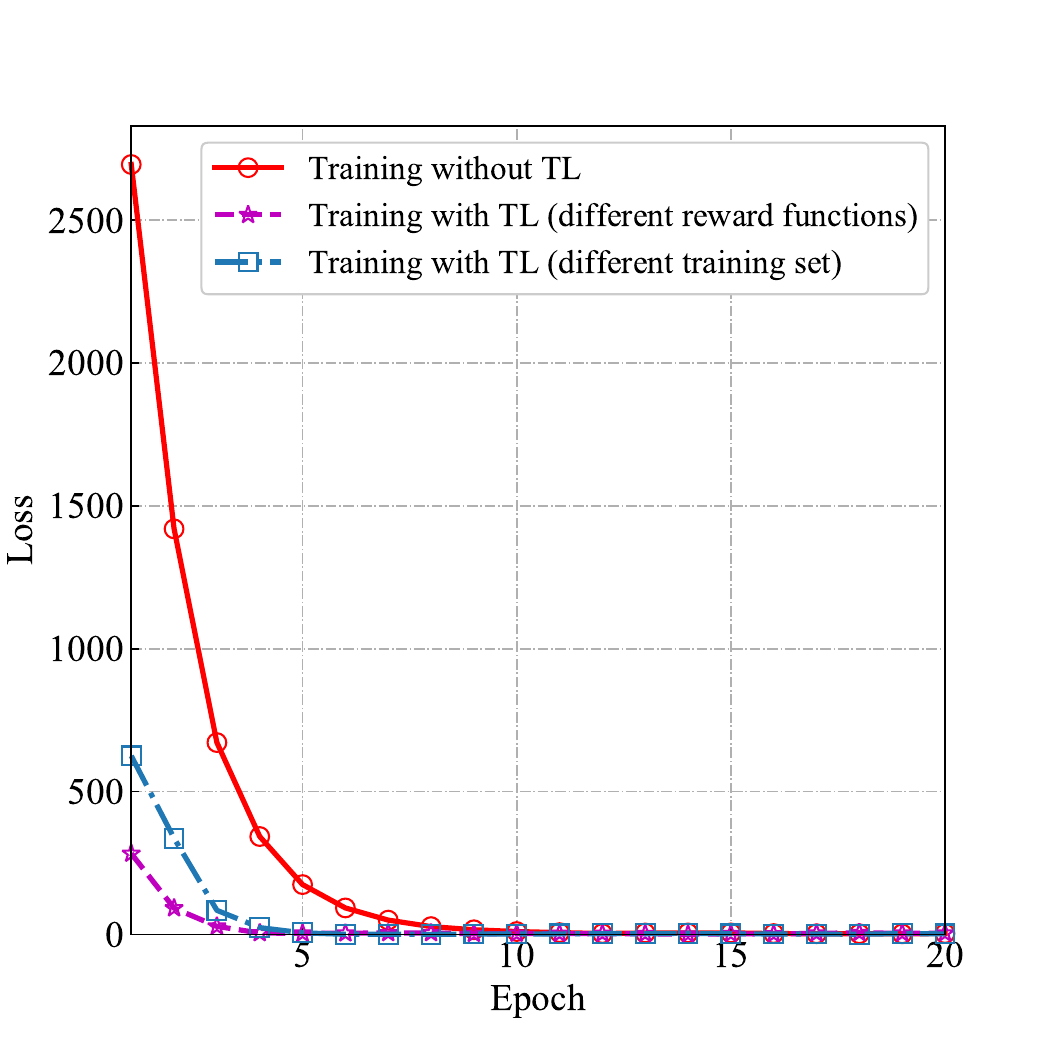}
		\caption{Loss values of transfer learning aided DPPD algorithm. }	
		\label{fig16}
		\vspace{-0.1cm}
\end{figure}

\subsection{Evaluation of Transfer Learning}
In this subsection, we use Scenario 2 as an example to present the performance of transfer learning using a small-scale network. For different reward functions, we use the pre-trained model from Scenario 1. For different training sets, we use the pre-trained model that was trained on the larger networks' training set.

\begin{figure}
\centering
\setlength{\abovecaptionskip}{0.cm}
\setlength{\belowcaptionskip}{-10 cm}
\includegraphics[width=6cm]{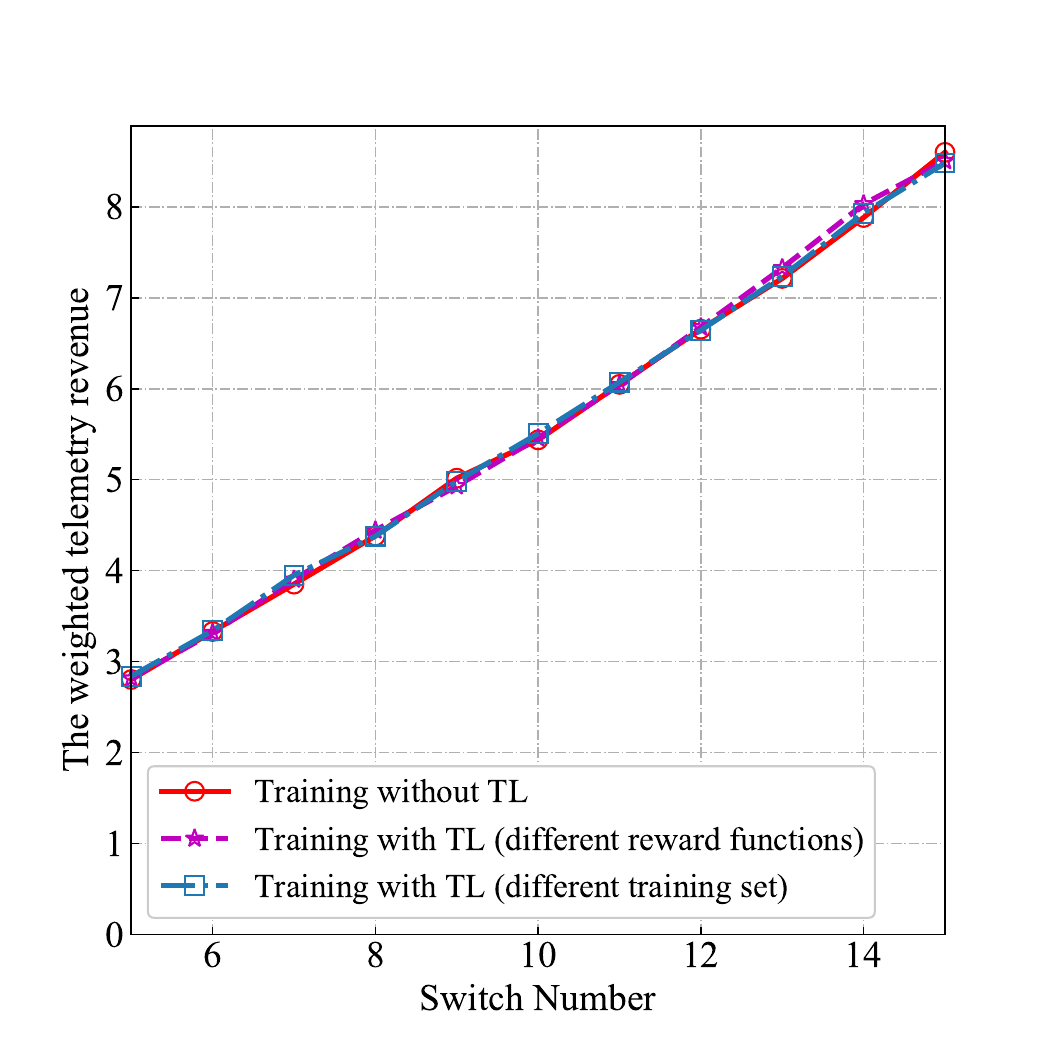}
\caption{The weighted telemetry revenue of transfer learning aided DPPD algorithm. }
\label{fig17}
\vspace{-0.1cm}
\end{figure}

Fig. \ref{fig16} shows the training efficiency of transfer learning. We observe that for both different reward functions and different training sets, transfer learning effectively reduces the training epochs needed to achieve convergence from approximately 9 to 4. Fig. \ref{fig17} shows that the performance of the path deployment using the training with transfer learning is similar to that without transfer learning. This indicates that transfer learning significantly reduces the complexity of training the DRL model and requires fewer epochs for model training without compromising the performance of the DPPD algorithm. This makes the DPPD algorithm superior in solving the challenge of telemetry requirements changes.

\vspace{0em}
\section{Conclusions}
\label{section7}
\vspace{-0.2em}
In this paper, we propose AdapINT, a dual-timescale network telemetry system. We design a network telemetry architecture consisting of APs and DPs, where APs are forwarded along auxiliary probe paths and DPs are forwarded along dynamic probe paths. We have developed two path deployment algorithms to make AdapINT adaptable to the dynamic network environment: a low-complexity APPD algorithm and a DRL-based DPPD algorithm. In addition, the DPPD algorithm also uses transfer learning to reduce the training time.
Simulation results show that AdapINT is capable of automatically identifying telemetry solutions that meet diverse telemetry requirements without the need for a manual calculation. Furthermore, our system demonstrates good robustness in dynamic network environments. In future work, we aim to address the challenge of reducing the complexity of model training for ultra-large-scale data center networks.

\bibliographystyle{ieeetr}\vspace{-0.3em}\vspace{-0.3em}% 规定参考文献的样式
\bibliography{INT_ref1}  %参考文献库的名字Ref

\begin{thebibliography}{10}

\bibitem{1}
L.~Tan, W.~Su, W.~Zhang, J.~Lv, Z.~Zhang, J.~Miao, X.~Liu, and N.~Li,
  ``{In-band network telemetry: A survey},'' {\em Computer Networks}, vol.~186,
  p.~107763, 2021.

\bibitem{6}
B.~Arzani, S.~Ciraci, L.~Chamon, Y.~Zhu, H.~H. Liu, J.~Padhye, B.~T. Loo, and
  G.~Outhred, ``{007: Democratically finding the cause of packet drops},'' in
  {\em 15th $\{$USENIX$\}$ Symposium on Networked Systems Design and
  Implementation ($\{$NSDI$\}$ 18)}, pp.~419--435, 2018.

\bibitem{7}
Y.~Li, R.~Miao, H.~H. Liu, Y.~Zhuang, F.~Feng, L.~Tang, Z.~Cao, M.~Zhang,
  F.~Kelly, M.~Alizadeh, {\em et~al.}, ``{HPCC: High precision congestion
  control},'' in {\em Proceedings of the ACM Special Interest Group on Data
  Communication}, pp.~44--58, 2019.

\bibitem{8}
P.~Tammana, R.~Agarwal, and M.~Lee, ``{Simplifying Datacenter Network Debugging
  with PathDump.},'' in {\em OSDI}, pp.~233--248, 2016.

\bibitem{9292999}
J.~Haxhibeqiri, P.~H. Isolani, J.~M. Marquez-Barja, I.~Moerman, and J.~Hoebeke,
  ``{In-Band Network Monitoring Technique to Support SDN-Based Wireless
  Networks},'' {\em IEEE Transactions on Network and Service Management},
  vol.~18, no.~1, pp.~627--641, 2021.

\bibitem{BHAMARE20171}
D.~Bhamare, M.~Samaka, A.~Erbad, R.~Jain, L.~Gupta, and H.~A. Chan, ``{Optimal
  virtual network function placement in multi-cloud service function chaining
  architecture},'' {\em Computer Communications}, vol.~102, pp.~1--16, 2017.

\bibitem{BHAMARE2016138}
D.~Bhamare, R.~Jain, M.~Samaka, and A.~Erbad, ``{A survey on service function
  chaining},'' {\em Journal of Network and Computer Applications}, vol.~75,
  pp.~138--155, 2016.

\bibitem{10063936}
B.~Chen, F.~Chen, S.~Tang, Q.~Zheng, and Z.~Zhu, ``{On Orchestration of Segment
  Routing and In-band Network Telemetry},'' {\em IEEE Transactions on Network
  and Service Management}, pp.~1--1, 2023.

\bibitem{3}
D.~Shan, F.~Ren, P.~Cheng, and R.~Shu, ``{Micro-burst in data centers:
  Observations, implications, and applications},'' {\em arXiv preprint
  arXiv:1604.07621}, 2016.

\bibitem{4}
T.~Benson, A.~Akella, and D.~A. Maltz, ``{Network traffic characteristics of
  data centers in the wild},'' in {\em Proceedings of the 10th ACM SIGCOMM
  conference on Internet measurement}, pp.~267--280, 2010.

\bibitem{2}
K.~Wu, J.~Xiao, and L.~M. Ni, ``{Rethinking the architecture design of data
  center networks},'' {\em Frontiers of Computer Science}, vol.~6,
  pp.~596--603, 2012.

\bibitem{9784426}
X.~Cheng, Z.~Wang, S.~Zhang, X.~He, and J.~J. Yang, ``{IntStream: Towards
  Flexible, Expressive, and Scalable Network Telemetry},'' {\em IEEE
  Transactions on Network and Service Management}, vol.~19, no.~3,
  pp.~2854--2868, 2022.

\bibitem{11}
C.~Kim, A.~Sivaraman, N.~Katta, A.~Bas, A.~Dixit, and L.~J. Wobker, ``{In-band
  network telemetry via programmable dataplanes},'' in {\em ACM SIGCOMM},
  vol.~15, 2015.

\bibitem{8897503}
S.~Tang, D.~Li, B.~Niu, J.~Peng, and Z.~Zhu, ``{Sel-INT: A Runtime-Programmable
  Selective In-Band Network Telemetry System},'' {\em IEEE Transactions on
  Network and Service Management}, vol.~17, no.~2, pp.~708--721, 2020.

\bibitem{9}
R.~Ben~Basat, S.~Ramanathan, Y.~Li, G.~Antichi, M.~Yu, and M.~Mitzenmacher,
  ``{PINT: Probabilistic in-band network telemetry},'' in {\em Proceedings of
  the Annual conference of the ACM Special Interest Group on Data Communication
  on the applications, technologies, architectures, and protocols for computer
  communication}, pp.~662--680, 2020.

\bibitem{10}
S.~Tang, J.~Kong, B.~Niu, and Z.~Zhu, ``{Programmable multilayer INT: An
  enabler for AI-assisted network automation},'' {\em IEEE Communications
  Magazine}, vol.~58, no.~1, pp.~26--32, 2020.

\bibitem{18}
E.~Song, T.~Pan, C.~Jia, W.~Cao, J.~Zhang, T.~Huang, and Y.~Liu, ``{INT-label:
  Lightweight in-band network-wide telemetry via interval-based distributed
  labelling},'' in {\em IEEE INFOCOM 2021-IEEE Conference on Computer
  Communications}, pp.~1--10, IEEE, 2021.

\bibitem{12}
T.~Pan, E.~Song, Z.~Bian, X.~Lin, X.~Peng, J.~Zhang, T.~Huang, B.~Liu, and
  Y.~Liu, ``{INT-path: Towards optimal path planning for in-band network-wide
  telemetry},'' in {\em IEEE INFOCOM 2019-IEEE Conference On Computer
  Communications}, pp.~487--495, IEEE, 2019.

\bibitem{13}
Y.~Lin, Y.~Zhou, Z.~Liu, K.~Liu, Y.~Wang, M.~Xu, J.~Bi, Y.~Liu, and J.~Wu,
  ``{Netview: Towards on-demand network-wide telemetry in the data center},''
  {\em Computer Networks}, vol.~180, p.~107386, 2020.

\bibitem{14}
D.~Bhamare, A.~Kassler, J.~Vestin, M.~A. Khoshkholghi, and J.~Taheri,
  ``{IntOpt: In-band Network Telemetry Optimization for NFV Service Chain
  Monitoring},'' in {\em ICC 2019-2019 IEEE International Conference on
  Communications (ICC)}, pp.~1--7, IEEE, 2019.

\bibitem{19}
Y.~Zhu, N.~Kang, J.~Cao, A.~Greenberg, G.~Lu, R.~Mahajan, D.~Maltz, L.~Yuan,
  M.~Zhang, B.~Y. Zhao, {\em et~al.}, ``{Packet-level telemetry in large
  datacenter networks},'' in {\em Proceedings of the 2015 ACM Conference on
  Special Interest Group on Data Communication}, pp.~479--491, 2015.

\bibitem{26}
M.~Nazari, A.~Oroojlooy, L.~Snyder, and M.~Tak{\'a}c, ``{Reinforcement learning
  for solving the vehicle routing problem},'' {\em Advances in neural
  information processing systems}, vol.~31, 2018.

\bibitem{TL}
H.~Xie, Z.~Qin, G.~Y. Li, and B.-H. Juang, ``{Deep Learning Enabled Semantic
  Communication Systems},'' {\em IEEE Transactions on Signal Processing},
  vol.~69, pp.~2663--2675, 2021.

\bibitem{SDN}
V.~Patel and D.~Vashi, ``{A Survey of Software-Defined Networking},'' 2017.

\bibitem{PDP}
R.~Bifulco and G.~Rétvári, ``{A Survey on the Programmable Data Plane:
  Abstractions, Architectures, and Open Problems},'' in {\em IEEE International
  Conference on High Performance Switching and Routing}.

\bibitem{ML1}
H.~Yao, T.~Mai, X.~Xu, P.~Zhang, M.~Li, and Y.~Liu, ``{NetworkAI: An
  Intelligent Network Architecture for Self-Learning Control Strategies in
  Software Defined Networks},'' {\em IEEE Internet of Things Journal}, vol.~5,
  no.~6, pp.~4319--4327, 2018.

\bibitem{28}
C.~A. Sunshine, ``{Source Routing in Computer Networks},'' {\em SIGCOMM Comput.
  Commun. Rev.}, vol.~7, p.~29–33, jan 1977.

\bibitem{5}
A.~Roy, H.~Zeng, J.~Bagga, G.~Porter, and A.~C. Snoeren, ``{Inside the Social
  Network's (Datacenter) Network},'' in {\em Proceedings of the 2015 ACM
  Conference on Special Interest Group on Data Communication}, SIGCOMM '15,
  (New York, NY, USA), p.~123–137, Association for Computing Machinery, 2015.

\bibitem{20}
V.~Mnih, K.~Kavukcuoglu, D.~Silver, A.~A. Rusu, J.~Veness, M.~G. Bellemare,
  A.~Graves, M.~Riedmiller, A.~K. Fidjeland, G.~Ostrovski, {\em et~al.},
  ``{Human-level control through deep reinforcement learning},'' {\em nature},
  vol.~518, no.~7540, pp.~529--533, 2015.

\bibitem{21}
H.~Van~Hasselt, A.~Guez, and D.~Silver, ``{Deep reinforcement learning with
  double Q-learning},'' in {\em Proceedings of the AAAI conference on
  artificial intelligence}, vol.~30, 2016.

\bibitem{23}
R.~S. Sutton and A.~G. Barto, {\em {Reinforcement learning: An introduction}}.
\newblock MIT press, 2018.

\bibitem{24}
J.~Schulman, S.~Levine, P.~Abbeel, M.~Jordan, and P.~Moritz, ``{Trust region
  policy optimization},'' in {\em International conference on machine
  learning}, pp.~1889--1897, PMLR, 2015.

\bibitem{25}
V.~Mnih, A.~P. Badia, M.~Mirza, A.~Graves, T.~Lillicrap, T.~Harley, D.~Silver,
  and K.~Kavukcuoglu, ``{Asynchronous methods for deep reinforcement
  learning},'' in {\em International conference on machine learning},
  pp.~1928--1937, PMLR, 2016.

\bibitem{9040280}
K.~Li, T.~Zhang, and R.~Wang, ``{Deep Reinforcement Learning for Multiobjective
  Optimization},'' {\em IEEE Transactions on Cybernetics}, vol.~51, no.~6,
  pp.~3103--3114, 2021.

\bibitem{16}
S.~Xiao, H.~Mao, B.~Wu, W.~Liu, and F.~Li, ``{Neural packet routing},'' in {\em
  Proceedings of the Workshop on Network Meets AI \& ML}, pp.~28--34, 2020.

\bibitem{15}
J.~Zhou, G.~Cui, S.~Hu, Z.~Zhang, C.~Yang, Z.~Liu, L.~Wang, C.~Li, and M.~Sun,
  ``{Graph neural networks: A review of methods and applications},'' {\em AI
  open}, vol.~1, pp.~57--81, 2020.

\bibitem{ML2}
R.~Hohemberger, A.~F. Lorenzon, F.~Rossi, and M.~C. Luizelli, ``{Optimizing
  Distributed Network Monitoring for NFV Service Chains},'' {\em IEEE
  Communications Letters}, vol.~23, no.~8, pp.~1332--1336, 2019.

\bibitem{sutskever2014sequence}
I.~Sutskever, O.~Vinyals, and Q.~V. Le, ``{Sequence to sequence learning with
  neural networks},'' {\em Advances in neural information processing systems},
  vol.~27, 2014.

\bibitem{vinyals2015pointer}
O.~Vinyals, M.~Fortunato, and N.~Jaitly, ``{Pointer networks},'' {\em Advances
  in neural information processing systems}, vol.~28, 2015.

\bibitem{bello2017neural}
I.~Bello, H.~Pham, Q.~V. Le, M.~Norouzi, and S.~Bengio, ``{Neural Combinatorial
  Optimization with Reinforcement Learning},'' 2017.

\bibitem{27}
A.~Roy, K.~Yocum, and A.~C. Snoeren, ``{Challenges in the emulation of large
  scale software defined networks},'' in {\em Proceedings of the 4th
  Asia-Pacific Workshop on Systems}, pp.~1--6, 2013.

\bibitem{el2021evaluating}
A.~El-Mekkawi, X.~Hesselbach, and J.~R. Piney, ``{Evaluating the impact of
  delay constraints in network services for intelligent network slicing based
  on SKM model},'' {\em Journal of Communications and Networks}, vol.~23,
  no.~4, pp.~281--298, 2021.

\end{thebibliography}
\bibliographystyle{unsrt}

\end{document}